\documentclass[aps,showpacs,letterpaper,twocolumn,12point]{revtex4-1}
\usepackage{graphicx} 
\usepackage{amsmath}
\usepackage{color}      

\usepackage{amstext}
\usepackage{graphics}
\usepackage{amssymb}

\newcommand{\ket}[1]{|#1\rangle}

\begin{document}

\title{Fidelity of a Rydberg blockade quantum gate from simulated quantum process tomography}

\author{X. L. Zhang, A. T. Gill, L. Isenhower,  T. G. Walker, and M. Saffman}

\affiliation{Department of Physics, University of Wisconsin, 1150 University Avenue,
Madison, WI 53706}

\begin{abstract}
We present a detailed error analysis of a Rydberg blockade mediated
controlled-NOT quantum gate between two  neutral
atoms as demonstrated recently in Phys. Rev. Lett. \textbf{104}, 010503
(2010) and Phys. Rev. A \textbf{82}, 030306 (2010).  
Numerical solutions of a master equation for the gate dynamics, including all known 
sources of technical error,  are shown to be in good agreement with experiments. 
The primary sources of gate error are identified and suggestions given for future improvements. 
We also present numerical simulations of quantum process tomography  to find the intrinsic fidelity, neglecting technical errors, 
of a Rydberg blockade controlled phase gate.
The gate fidelity is characterized using trace overlap and trace distance measures. We show that the trace distance is 
linearly sensitive to errors arising from the finite Rydberg blockade shift and introduce a modified pulse sequence which corrects the 
linear errors. 
Our  analysis shows that the intrinsic gate error extracted from simulated quantum process tomography can be under 0.002 for  specific states of $^{87}$Rb or Cs atoms.
The relation between the process fidelity and the gate error probability used in calculations of fault tolerance thresholds is discussed. 
\end{abstract}

\pacs{03.67.-a, 32.80.Qk, 32.80.Ee.}

\maketitle

\section{Introduction}

Arrays of laser cooled neutral atoms in optical lattices or far-off-resonance optical traps (FORTs) are  promising candidates 
for quantum computing experiments, due to their very long decoherence time (up to several seconds) 
for ground state atoms and strong two-atom interaction for highly excited Rydberg state atoms. 
This strong, long-range and controllable interaction leads to the so-called Rydberg blockade effect in which  only one atom in an ensemble can be excited into a Rydberg state 
if the ensemble size is smaller than  the Rydberg blockade radius.
Jaksch et al. \cite{Jaksch2000} first proposed to use Rydberg blockade to implement 
a fast two-qubit controlled-phase gate (C$_Z$), which can be converted into a controlled-NOT gate (CNOT)
using single qubit rotations \cite{Nielsen2000}. 
Soon after, various schemes were proposed for fast quantum gates with
an ensemble \cite{Lukin2001,Saffman2005b,Isenhower2011,H-ZWu2010},
entangled state preparation \cite{Moller2008,*Muller2009,*Saffman2009b}, quantum algorithms \cite{AChen2011,Molmer2011}, 
quantum simulators \cite{Weimer2010,*Weimer2011},
and efficient quantum repeaters \cite{Han2010,*Zhao2010}.

Rydberg blockade, the central ingredient of the above schemes, has been demonstrated 
 between two individual neutral atoms held in FORTs \cite{Urban2009,Gaetan2009}, 
and was used to demonstrate a two-qubit CNOT quantum gate \cite{Isenhower2010} 
and to generate entangled Bell states with fidelity of about $0.58-0.75$ after atom loss correction ($0.46-0.48$
without atom loss correction) \cite{Isenhower2010,Wilk2010,Gaetan2010}.
Using an improved apparatus \cite{Zhang2010} a fidelity of $0.92$ for the CNOT probability truth table
and $0.71$ for Bell state fidelity after atom loss correction ($0.71/0.58$ for the CNOT truth table/Bell state fidelity 
without atom loss correction) was demonstrated. 
These proof-of-principle results are promising but are still far from simple estimates of  Rydberg gate errors 
at the level of $E\sim10^{-3}$ predicted in \cite{Jaksch2000,Saffman2005a}. 

The simple estimates are based on intrinsic errors associated with the atomic physics of the states used for Rydberg blockade. 
The essential intrinsic errors are the finite lifetime of Rydberg states and the finite strength of the Rydberg-Rydberg blockade interaction. 
In addition to intrinsic errors, experiments are sensitive to  several different sources of technical error:  
spontaneous emission from an intermediate level in a two-photon excitation scheme, magnetic field fluctuations, 
pulse area errors, Doppler effects due to finite atomic temperature, etc.. 
We  have previously estimated gate errors \cite{Saffman2005a,Zhang2010} by treating each error source separately
 in the small error limit and adding them together.
This provides a good estimate for small errors but it is unreliable for larger errors, as in the experiments,
and does not provide a rigorous fidelity measure for the gate operation. 

In this paper we present a more rigorous treatment by including all known error sources in a master equation
 for the density matrix evolution (optical Bloch equations). 
We  dynamically track the density matrix evolution during the  CNOT pulse sequence and then average the results 
over the computational input states for  comparison with simple analytical estimates. We also extract a rigorous  value for the gate 
fidelity using simulated quantum process tomography. 
The numerical results are in good agreement with  analytical error estimates when technical errors are neglected, and with experimental results when technical errors are included in a Monte Carlo simulation. 
Numerical simulations of quantum process tomography with realistic atomic  parameters 
 confirm that it is in principle possible to reach quantum process errors of $2\times 10^{-3}$ for both Rb and Cs atoms.

In Sec. \ref{sec.experiment} we present the experimental setup and procedures used to demonstrate the CNOT gate and generate entangled Bell states. We also enumerate the various sources of technical imperfection in a realistic experiment.  
In Sec. \ref{sec.errors} we give analytical estimates of the intrinsic gate error and present a new pulse sequence which removes the leading linear term in the finite blockade shift error. 
In Sec. \ref{sec.master} we present a master equation model which includes  both the technical error sources from \ref{sec.experiment} and intrinsic errors from Sec. \ref{sec.errors}. In Sec. \ref{sec.masterexp} we compare numerical Monte Carlo simulations with experimental results, and demonstrate good agreement.
In Sec. \ref{sec.masterlimit} we perform simulated quantum process tomography of a two qubit controlled-phase gate accounting only for  intrinsic errors. This analysis shows that in a well designed experiment where technical errors are minimized it should be possible to reach low gate errors, below known fault tolerance thresholds\cite{Knill2005,Raussendorf2007,*Aliferis2009,*Wang2011},  for both Rb and Cs. 
A summary and discussion is presented in Sec. \ref{sec.discussion}.

\section{experimental procedures and technical imperfections}
\label{sec.experiment}

The experimental apparatus and procedures, as shown in Fig. \ref{fig:setup},
have been described in detail in our recent publications \cite{Isenhower2010,Zhang2010}.
For ease in understanding the subsequent analysis we include a brief description of the procedures, as well as estimates of various sources of technical imperfections.

\subsection{Experimental procedures}

\begin{figure}[!t]
\begin{centering}
\includegraphics[width=1\columnwidth]{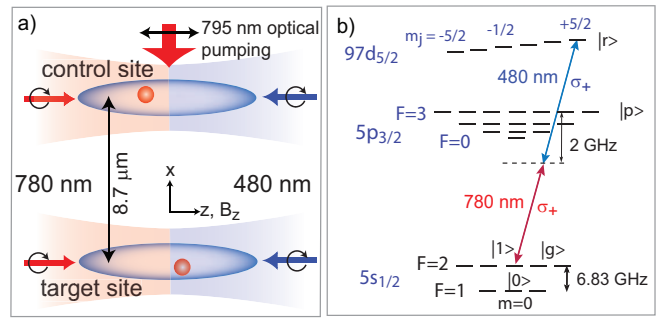}
\par\end{centering}
\caption{(color online) a) Experimental geometry. Two single atoms are trapped in two FORTs separated by about $9~\mu \rm m$.
 b) Relevant levels of $^{87}$Rb. To reach the Rydberg state we use two-photon excitation 
with wavelengths of $780~\rm nm$ and $480~\rm nm$.\label{fig:setup}}
\end{figure}

We use FORTs to localize single $^{87}$Rb atoms, which can be individually addressed using tightly focused beams 
that are scanned by acousto-optical modulators. 
The FORT beams, propagating along $+z$, are formed by  focusing a laser beam with wavelength of 
$\lambda_{\rm f}=1064~\rm nm$ to a waist ($1/e^2$ intensity radius) of $w_{\rm f}=3.4~\mu \rm m$. 
We generate a linear array of  trap sites using a diffractive element with the central site's trap
depth of $U_{0}/k_{B}=4.5~{\rm mK}$ and trap separation of about $9~\mu{\rm m}$ along the $x$ direction. 
We  use two sites, one labeled as the control  and the other as the target,
to perform two-qubit quantum logic operations and to generate entangled states. 
A bias magnetic field is applied along $z$, which defines the quantization axis for the optical pumping
 ($B_{z}=0.15 {\rm ~mT}$) and lifts the degeneracy of the Zeeman sublevels ($B_{z}= 0.37{\rm ~mT}$)
of Rydberg states during the gate operation. 
The relevant levels of $^{87}$Rb are shown in Fig. \ref{fig:setup}b. 
We use the $5s_{1/2}$ hyperfine clock states as our qubits $|0\rangle\equiv|F=1,m_{F}=0\rangle$
and $\mbox{\mbox{\ensuremath{|1\rangle\equiv|F=2,m_{F}=0\rangle}}}$,
separated by $\omega_{10}/2\pi=\mbox{6.83{\rm ~GHz}}$, 
and the Rydberg state $|r\rangle \equiv|97d_{5/2},m_{j}=5/2\rangle$.

We perform single qubit rotations between $|0\rangle$ and $|1\rangle$ using two-photon stimulated Raman
transitions driven by focusing a $\sigma_{+}$ polarized $780~\rm nm$ laser with frequency components 
separated by $\omega_{10}$ and detuned by $101{\rm ~GHz}$ to the red of the D2 transition \cite{Yavuz2006}.
Total typical power in the two Raman sidebands is $\sim90{\rm ~}\mu{\rm W}$ with waist of
 $\omega_{x/y,g}=7.7{\rm ~}\mu{\rm m}$, giving a  single qubit Rabi frequency of 
$\Omega_{g}/2\pi=0.56{\rm ~MHz}$ with $\pi$ pulse times of $\sim900~\rm ns$ and peak-to-peak amplitude
of better than 0.98 after correction for background atom loss of $10\%$ \cite{Zhang2010}.

For coherent Rydberg excitation between $|1\rangle$ and $|r\rangle$ we use a two-photon transition
with $\sigma_{+}$ polarized 780 and $480~\rm nm$ beams \cite{Johnson2008}.
They counter-propagate along the trap's axial $z$ direction to minimize Doppler broadening of the transition. 
The $780~\rm nm$ beam is tuned about $2.0{\rm ~GHz}$ to the red of
the $\ket{1}\rightarrow|5p_{3/2},F^{'}=3\rangle$ transition
with typical beam power of $2.4{\rm ~}\mu{\rm W}$ and waist of $\omega_{x/y,R}=7.7{\rm ~}\mu{\rm m}$.
The $480~\rm nm$ beam is tuned about $2.0{\rm ~GHz}$ to the blue of
 the $|5p_{3/2},F^{'}=3\rangle\rightarrow\ket{r}$ transition with typical beam power of $13~\rm mW$ 
and waist of $\omega_{x/y,B}=4.5{\rm ~}\mu{\rm m}$. 
This gives a Rydberg red Rabi frequency of $\Omega_{\rm R}/2\pi=118{\rm ~MHz}$, 
a Rydberg blue Rabi frequency of $\Omega_{\rm B}/2\pi=39{\rm ~MHz}$, 
and Rydberg Rabi frequency of $\Omega/2\pi=1.15{\rm ~MHz}$ with $\pi$ pulse times
of $\sim 440~\rm ns$ and amplitude of 0.92 after correction for background atom loss of $10\%$.

In order to perform a two qubit CNOT gate we start by loading single atoms 
from a background vapor magneto-optical trap into two FORT sites.
The trapped atoms have a measured temperature of $T_{\rm a}\simeq175~\mu{\rm K}$
using a release and recapture method \cite{Reymond2003,*Tuchendler2008}.
Atom detection is accomplished by simultaneously illuminating both trapping sites with near resonant red-detuned
molasses light, and imaging the fluorescence onto a cooled EMCCD camera.
Detected photon counts are integrated for approximately 20 ms. 
Comparison of  the integrated number of counts in a region of interest with predetermined thresholds 
indicates the presence or absence of a single atom \cite{Urban2009}. After single atoms are loaded
in these sites they are optically pumped into $|1\rangle$  with efficiency of about $99\%$ \cite{Zhang2010}
using $\pi$ polarized light propagating 
along $-x$ tuned to the $|5s_{1/2},F=2\rangle\rightarrow|5p_{1/2},F^{'}=2\rangle$ D1 transition
at $795~\rm nm$ and $|5s_{1/2},F=1\rangle\rightarrow|5p_{3/2},F^{'}=2\rangle$
D2 transition at $\mbox{780~nm}$. 
This is followed by ground state $\pi$ pulses to either or both of the atoms to generate any of 
the four computational basis states. 
We then turn off the optical trapping potentials for about $4~\mu\rm s$, apply the CNOT pulses, and restore the optical traps.
Ground state $\pi$ pulses are then applied to either or both atoms to select one of the
four possible output states. 
Atoms left in state $|1\rangle$ are removed from the traps with unbalanced radiation pressure 
(blow away light), and a measurement is made to determine if the selected output state is present. 
The selection pulses provide a positive identification of all output states and 
we do not simply assume that a low photoelectron signal corresponds to an atom in $|1\rangle$ 
before application of the blow away light \cite{Isenhower2010,Zhang2010}.

Following the above procedures we have obtained the CNOT truth table fidelity of
 $F=\frac{1}{4}{\rm Tr}[U_{{\rm ideal}}^{T}U_{{\rm CNOT}}]=0.92\pm0.06$
with $U_{\rm ideal}$ and $U_{\rm CNOT}$ the ideal and experimentally obtained CNOT truth tables.
To measure the state preparation fidelity, we use the same sequence but without applying the CNOT pulses. 
The computational basis states are prepared with an average fidelity of $F=0.97$.

To create entangled states we use $\pi/2$ pulses on the control atom to prepare the input states
 $|ct\rangle=\frac{1}{\sqrt{2}}(|0\rangle+i|1\rangle)|1\rangle$ and
 $|ct\rangle=\frac{1}{\sqrt{2}}(|0\rangle+i|1\rangle)|0\rangle$.
Applying the CNOT to these states creates two of the Bell states 
$|B_{1}\rangle=\frac{1}{\sqrt{2}}(|00\rangle+|11\rangle)$ and 
$|B_{2}\rangle=\frac{1}{\sqrt{2}}(|01\rangle+|10\rangle$). 
In order to verify entanglement we measured the coherence of the $|B_{1}\rangle$ state by parity oscillations \cite{Isenhower2010,Zhang2010}, and obtained a Bell state fidelity of $0.58$ 
without any atom loss correction ($0.71$ after  correction) \cite{Zhang2010}. 
Comparable numbers for the Bell state fidelity were obtained in a related experiment \cite{Wilk2010}.

\subsection{Technical imperfections}
\label{sec.technical}

The main technical errors that affect the CNOT operation are spontaneous emission of the intermediate level during Rydberg excitation, magnetic noise, Doppler effects due to finite atom temperature and Rydberg laser power fluctuations.

The spontaneous emission error of the intermediate level during a $\pi$ excitation pulse can be estimated by 
$P_{\rm se}=\frac{\pi\gamma_{p}}{4|\Delta_{p}|}(|\frac{\Omega_{\rm R}}{\Omega_{\rm B}}|+|\frac{\Omega_{\rm B}}{\Omega_{\rm R}}|)$,
with $\gamma_p, \Delta_p$ the radiative linewidth and detuning from the $p$ level. 
This error is about $0.8\%$ for our current experimental setup.

Magnetic field fluctuations cause transition shifts, giving a  Rydberg two-photon detuning 
$\Delta_{\rm B}=(g_{r}m_{r}-g_{1}m_{1})\mu_{B}B_{z}$
with $g_{r}= 6/5$, $m_{r}=5/2$, $g_{1}=1/2$, and $m_{1}=0$ 
for our implementation in Fig. \ref{fig:setup}b). 
We assume that the magnetic field fluctuations are Gaussian distributed with a standard
deviation of $\sigma_{\rm B}=2.5\times10^{-6}~T$. 
This value was found  by measuring the decoherence time of the hyperfine qubit 
at two different bias magnetic field strengths\cite{Saffman2011}.
Doppler broadening at finite atom temperature $T_{\rm a}$,
 also gives two-photon detuning $\Delta_{\rm D}=(k_{R}-k_{B})v$ with averaged variances of $\sigma_{vx}^{2}=\sigma_{vy}^{2}=\sigma_{vz}^{2}=\frac{k_B T_{\rm a}}{m}$  for both control and target atoms.

\begin{figure}
\begin{centering}
\includegraphics[width=1\columnwidth]{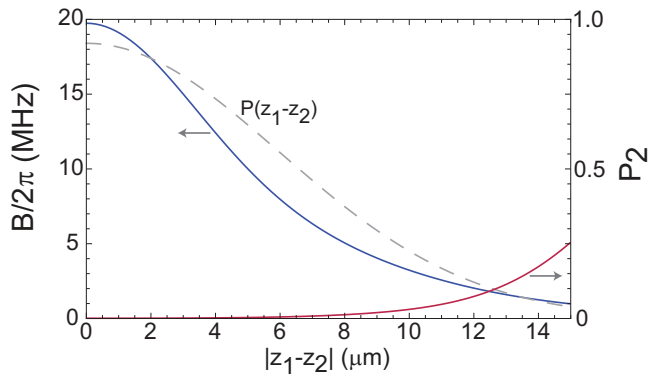}
\par\end{centering}
\caption{(color online) Calculated blockade shift $\sf B$ from the theory of Ref. \cite{Walker2008} and double excitation probability $P_{2}$ for state $|97d_{5/2},m_{j}=5/2\rangle$ 
as a function of relative position $P(|z_{1}-z_{2}|)$ for $0.37~\rm mT$ magnetic field.
The relative probability distribution $P(|z_{1}-z_{2}|)$ assumes a trapping potential $U/k_{B}=4.5~{\rm mK}$, 
waist $w=3.4~\mu{\rm m}$ of the 1064 nm trapping light and atom temperature T$=175~\mu{\rm K}$.
\label{fig:blockadeshift}}
\end{figure}

Other technical errors associated with finite atom temperature are Rydberg blockade shifts, pulse area fluctuations, 
and AC Stark shifts (or two-photon detunings) due to the atomic position distribution in the FORTs. 
We assume that the atomic position distribution for both control and target atoms is Gaussian 
with variance of \cite{Saffman2005a} $\sigma_{x}^{2}=\sigma_{y}^{2}=\frac{w_{f}^{2}}{4}\frac{k_{B}T_{\rm a}}{U_{0}}$
and $\sigma_{z}^{2}=\frac{\pi{}^{2}w_{\rm f}^{4}}{2\lambda_{\rm f}^{2}}\frac{k_{B}T_{\rm a}}{U_{0}}$.
For the Rydberg blockade shift due to the atomic position distribution,
we use the theoretical blockade shift curve as a function of relative atomic separation
 $|z_{1}-z_{2}|$ as shown in Fig. \ref{fig:blockadeshift}.
The position dependent Rabi frequency and AC Stark shifts are 
\begin{eqnarray}
\Omega_{\rm R}({\bf r})&=&\Omega_{\rm R}(0)\frac{e^{-\left[\frac{x^{2}}{w_{x,R}^{2}\left(1+z^{2}/L_{x,R}^{2}\right)}+\frac{y^{2}}{w_{y,R}^{2}\left(1+z^{2}/L_{y,R}^{2}\right)}\right]}}{[(1+z^{2}/L_{x,R}^{2})(1+z^{2}/L_{y,R}^{2})]^{1/4}}\nonumber\\
\\
\Omega_{\rm B}({\bf r})&=&\Omega_{\rm B}(0)\frac{e^{-\left[\frac{x^{2}}{w_{x,B}^{2}\left(1+z^{2}/L_{x,B}^{2}\right)}-\frac{y^{2}}{w_{y,B}^{2}\left(1+z^{2}/L_{y,B}^{2}\right)}\right]}}{[(1+z^{2}/L_{x,B}^{2})(1+z^{2}/L_{y,B}^{2})]^{1/4}}\nonumber\\
\end{eqnarray}
and
\begin{equation}
\Delta_{\rm AC}({\bf r})=\Delta_{\rm AC}(0)+\frac{\Omega_{\rm R}^{2}({\bf r})-\Omega_{\rm B}^{2}({\bf r})}{4|\Delta_{p}|}.
\end{equation}
Here $\Omega_{\rm R}(0)$, $\Omega_{\rm B}(0)$, and $\Delta_{\rm AC}(0)$ are 
the Rydberg red, blue and AC Stark shifts at trap center, respectively;
the Rayleigh lengths are $L_{x/y,R}=\pi w_{x/y,R}^{2}/\lambda_{\rm R}$
and $L_{x/y,B}=\pi w_{x/y,B}^{2}/\lambda_{\rm B}$.

Power fluctuations of the Rydberg lasers will not only affect the Rabi frequencies $\Omega_{\rm R}$ and $\Omega_{\rm B}$, 
but will also affect the two photon detuning $\Delta_{\rm AC}$.
We assume that the power fluctuations of the red and blue lasers are both Gaussian distributed with 
FWHM of $1\%$ and $2\%$, respectively, as measured independently.

\subsection{Dephasing errors}

The technical imperfections listed in the previous section show up as errors in the measured CNOT probability truth table as well as in the fidelity of the output quantum states. There are additional technical dephasing errors that do not significantly affect the CNOT truth table but strongly 
impact the fidelity of Bell state generation. 
As was pointed out in Refs. \cite{Wilk2010,Saffman2011}  both magnetic field fluctuations
 and atomic motion
lead to dephasing of the Rydberg state relative to the ground state during gate operation
because the motion of Rydberg atoms between excitation and deexcitation pulses leads to 
a stochastic phase that degrades the entanglement.
In the numerical simulations described in  Sec. \ref{sec.master}  we do not keep track of the 
position dependent phase of the optical fields  in the evolution Hamiltonians
during each blockade pulse sequence(Eqs. (\ref{eq:Hamiltonian}) and (\ref{eq:Hamiltonian2}) below) .
Instead we add an extra dephasing term  $\gamma_{\rm ph}$ to the Liouville  operators of 
Eqs. (\ref{eq:Liouville}) and (\ref{eq:Liouville2}) in Sec. \ref{sec.master}:
\begin{equation}
\gamma_{\rm ph}=\sqrt{(\gamma_{\rm B})^{2}+(\gamma_{\rm D})^{2}},
\label{eq:PhaseError} 
\end{equation}
where  $\gamma_{\rm B}=|\Delta_{\rm B}|/\hbar$
and $\gamma_{\rm D}=\Delta_{\rm D}$ are the dephasing rates due to magnetic field fluctuations and Doppler effects, respectively.
In the Monte Carlo simulations presented below these dephasing rates are sampled from distributions that are generated with the position and velocity variances described above.
Both the magnetic field fluctuations and atom position variations at finite temperature 
also dephase the qubit states by varying the hyperfine splitting between them. 
We model the qubit dephasing as 
\begin{equation}
\gamma_{01}=\sqrt{(\gamma_{B01})^{2}+(\gamma_{T})^{2}}
\end{equation}
where $\gamma_{B01}=\omega_{10}(B_{z}+\sigma_{\rm B}) - \omega_{10}(B_{z})$, is the dephasing rate 
due to the second-order Zeeman shift of the clock transition by magnetic field fluctuations,  $B_z$ is a static bias field, 
$\omega_{10}(B_{z}) = \omega_{10} \sqrt{1+\left[\frac{(g_{S}-g_{I}) \mu_{B} B_{z}}{\hbar \omega_{10}}\right]^2} $
 is the ground hyperfine splitting of the clock states, $g_S,g_I$ are electron and nuclear Land\'e factors,  and $\sigma_{B}$ is the magnetic field fluctuation; $\gamma_{T}=1.03\frac{\delta_{0}k_{B}T_{\rm a}}{2U_{0}}$ \cite{Saffman2005a,Kuhr2005},
 is the dephasing rate in 1/ms due to atomic motion in the FORT, $U_{0}$ is the peak FORT potential,
$\delta_{0}/2\pi=(U_{0}/k_B)\times(1.5~\rm kHz/mK)$, is the peak differential light shift of the FORT 
and $T_{\rm a}$ is the atom temperature.

In addition to the above  errors,  there are also  errors associated with state preparation due to imperfect 
optical pumping and single qubit rotations. These errors  are at about the  $1\%$ level. There is also  about $1\%$ atom loss due to background collisions before the CNOT pulse sequence. 

\section{Intrinsic error estimates}
\label{sec.errors}

Even if all sources of technical error described in Sec. \ref{sec.technical} are negligible there will still be intrinsic gate errors due to the basic physics of the Rydberg blockade interaction. 
Intrinsic errors of a Rydberg blockade CNOT gate include the decoherence error  due to 
the finite lifetime $\tau$ of the Rydberg state and rotation errors due to imperfect blockade.
In the strong blockade limit ($\Omega\ll {\sf B}\ll\omega_{10}$, where $\sf B$ is the Blockade shift), 
the intrinsic gate error $E_1$ averaged over the input states in the computational basis 
$(\{\ket{00},\ket{01},\ket{10},\ket{11}\})$  is \cite{Saffman2005a,Saffman2010}

\begin{equation}
E_1=\frac{7\pi}{4\Omega\tau}\left(1+\frac{\Omega^{2}}{\omega_{10}^{2}}+\frac{\Omega^{2}}{7{\sf B}^{2}}\right)+\frac{\Omega^{2}}{8{\sf B}^{2}}\left(1+6\frac{\Omega^{2}}{\omega_{10}^{2}}\right)
\label{eq:IntrinsicError1}
\end{equation}

The first term in Eq. (\ref{eq:IntrinsicError1}) is the Rydberg decay error 
due to the finite lifetime $\tau$ of the Rydberg state,  and the second term is the imperfect blockade error. 
In the limit of $\omega_{10}\gg({\sf B},\Omega)$ 
we can extract a simple expression for the optimum Rabi frequency which minimizes the error
\begin{equation}
\Omega_{1_{\rm opt}}=(7\pi)^{1/3}\frac{{\sf B}^{2/3}}{\tau^{1/3}}.
\label{eq:OptRabi1}
\end{equation}
Setting $\Omega\rightarrow\Omega_{1_{\rm opt}}$ leads to a minimum averaged gate error of
\begin{equation}
E_{1_{\rm min}}=\frac{3(7\pi)^{2/3}}{8}\frac{1}{({\sf B}\tau)^{2/3}}.
\label{eq:MinError1}
\end{equation}
In our experiments $\tau\sim300{\rm ~}\mu{\rm s}$ is the radiative lifetime of the $97d_{5/2}$ Rydberg level, $\omega_{10}/2\pi=6.83~{\rm GHz}$, and $\Omega/2\pi=1.15~{\rm MHz}$.
In the experimental geometry shown in Fig. \ref{fig:setup}a) a range of two-atom separations,
and hence blockade shifts, occur. 
The blockade shift curve shown in Fig. \ref{fig:blockadeshift} is calculated from the theory of Ref.
\cite{Walker2008} using a trap separation of $x=8.7~\mu{\rm m}$ and 
a bias magnetic field of $B_{z}=0.37~{\rm mT}$ applied along the $\hat{z}$ axis. 
Averaging Eq. (\ref{eq:IntrinsicError1}) over the probability distribution P$(|z_{1}-z_{2}|)$, 
which is dependent on the trapped atom temperature of $175~\mu{\rm K}$, gives 
an expected error of $E=8.5\times10^{-3}$. 
The corresponding averaged blockade shift from Eq. (\ref{eq:IntrinsicError1})  is $\bar{\sf B}/2\pi=5.3~{\rm MHz}$.

It should be emphasized that the average error $E_1$ discussed above ignores errors in the phases of the states 
generated by the CNOT gate and therefore corresponds to the measurement of a probability truth table. 
As discussed in Ref. \cite{Jaksch2000}, a Rydberg-mediated C$_Z$  gate has a phase error for the $\ket{11}$ input state of $\ket{11}\rightarrow -e^{\imath\phi}\ket{11}$ with $\phi=\pi \Omega/2{\sf B}$.
Averaging over the four computational basis states gives an average intrinsic phase error of $E_{\phi}=\pi\Omega/8{\sf B}$.
Including  the phase error, we find an average intrinsic gate error of
\begin{equation}
E_2=E_1+\frac{\pi\Omega}{8{\sf B}}.
\label{eq:IntrinsicError2}
\end{equation}
When  $\Omega\ll {\sf B}$ the last term in Eq. (\ref{eq:IntrinsicError2}), 
dominates over the imperfect blockade term in Eq. (\ref{eq:IntrinsicError1}).
Using Eq. (\ref{eq:IntrinsicError2}) we can again extract a simple expression for the optimum Rabi frequency which minimizes the error
\begin{equation}
\Omega_{2_{\rm opt}}=\left(\frac{14{\sf B}}{\tau}\right)^{1/2}.
\label{eq:OptRabi2}
\end{equation}
Setting $\Omega\rightarrow\Omega_{2_{\rm opt}}$ leads to a minimum averaged gate error of
\begin{equation}
E_{2_{\rm min}}=\frac{\sqrt{7}\pi}{2\sqrt{2}}\frac{1}{({\sf B}\tau)^{1/2}}.
\label{eq:MinError2}
\end{equation}
Comparing Eqs. (\ref{eq:MinError1}) and (\ref{eq:MinError2}) in the experimentally relevant limit of ${\sf B}\tau\gg 1$ 
we see that $E_{2_{\rm min}}\gg E_{1_{\rm min}}$. 
This result seems to imply that  Eq.  (\ref{eq:MinError1})  provides an overly optimistic estimate for the gate error. 

\begin{figure}
\begin{centering}
\includegraphics[width=1\columnwidth]{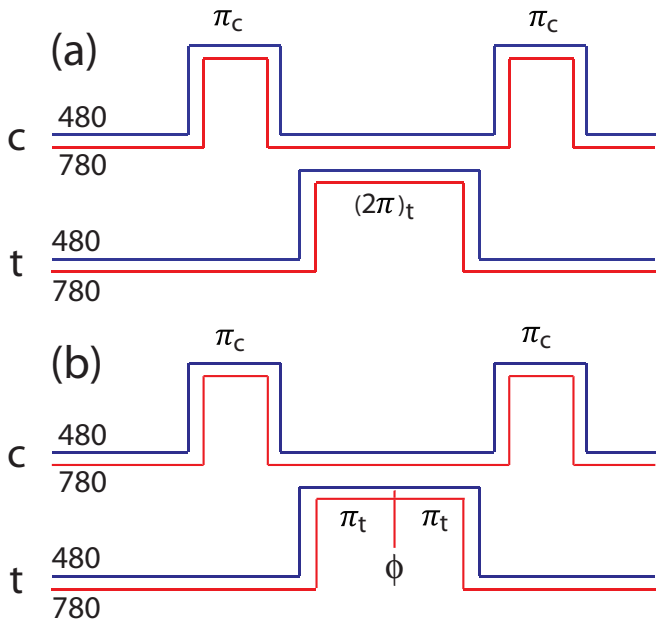}
\par\end{centering}
\caption{(color online) Pulse sequence for Rydberg C$_Z$ gate with two-photon 780 and 480 nm excitation of Rb atoms following\cite{Jaksch2000} a) and modified sequence to correct finite blockade phase error by rapidly changing the phase of one of the Rydberg beams by $\phi$ b). 
\label{fig:sequence}}
\end{figure}

Despite this apparently disappointing result the gate error can indeed be made to satisfy  the
$({\sf B}\tau)^{-2/3}$ scaling of Eq. (\ref{eq:MinError1}) by a  modification of the C$_Z$ sequence as shown in Fig. \ref{fig:sequence}b).
We note that the only input state for which the target atom becomes Rydberg excited  is $\ket{01}$. 
By shifting the phase of the  $\ket{5p_{3/2}}\rightarrow \ket{r}$ laser used for Rydberg excitation by an amount $\phi$ during the gate operation we can add a phase of $\phi$ to this state. 
This results in a phase gate 
\begin{equation}
C_Z  =  \left(\begin{array}{cccc}
1 & 0 & 0 & 0\\
0 & -e^{\imath\phi} & 0 & 0\\
0 & 0 & -1 & 0\\
0 & 0 & 0 & -e^{\imath\phi}\end{array}\right).
\label{eq:Cz2}
\end{equation}
Adding a single qubit $Z$ rotation of $-\phi$ to the control atom  cancels the extra phases, leaving an ideal C$_Z$ operation. 
As we show in detail in Sec. \ref{sec.masterlimit} using simulated process tomography, the standard sequence of Fig. \ref{fig:sequence}a gives 
linear phase errors that impact the trace distance fidelity measure. Using the modified pulse sequence the linear phase errors are canceled and the trace distance error is reduced.  

We emphasize that perfect correction of the phase error using this modified gate sequence assumes that ${\sf B}$, and therefore the phase $\phi$, are well defined, and do not fluctuate. This is not true for thermally excited atoms but will be the case for atoms that are in the motional ground state of the confining potentials. If the atoms are thermally excited we can still cancel the average value of $\phi$, but errors due to fluctuations about the average would remain. In the analysis of the intrinsic error limit we assume that the atoms are  in the motional ground state so the phase can be canceled exactly.

\section{numerical simulations}
\label{sec.master}

In order to accurately simulate the performance of the Rydberg gate we integrate the master equation for the two-atom dynamics including all known coherent and incoherent rates. A related analysis was performed previously for the Rydberg excitation dynamics of a single atom \cite{Miroshnychenko2010}.  
For each atom we include five atomic states in our numerical calculation, which are labeled 
in the level scheme of Fig. \ref{fig:setup}b): qubit $|0\rangle$, qubit $|1\rangle$, 
and reservoir level $|g\rangle\equiv|{5s}_{1/2},m_{F}\neq0\rangle$
in the ${5s}_{1/2}$ ground state; the intermediate level $|p\rangle\equiv|5p_{3/2}\rangle$ 
and the Rydberg level $|r\rangle$. With this set of basis states the two-atom dynamics are described by density matrices $\rho_{\rm ct}(t)$ with dimensions $25\times 25$. We take the initial condition to be a separable state $\rho_{\rm ct}(0)=\rho_{\rm c}(0)\otimes\rho_{\rm t}(0)$, with c/t for control/target atoms.
We calculate the time evolution by solving the master equation
\begin{equation}
\frac{d\rho_{\rm ct}}{dt}=-\frac{i}{\hbar}[H_{\rm ct},\rho_{\rm ct}]+\mathcal{L_{\textrm{ct}}},
\label{eq:master}
\end{equation}
with  
${H}_{\rm ct}={H}_{\rm c}\otimes I_{\rm t}+I_{\rm c}\otimes{H}_{\rm t}+\hbar{\sf B}\left[\begin{array}{cc}
0_{24} & 0\\0 & 1\end{array}\right]$, 
$\mathcal{L}_{\rm ct}=\mathcal{L}_{\rm c}\otimes I_{\rm t}+I_{\rm c}\otimes\mathcal{L}_{\rm t}$, $I_{\rm t}$ $(I_{\rm c})$ 
are  $5\times5$ identity matrices, and 0$_{24}$ is a  $24\times24$ zero matrix. The  Hamiltonian ${H}_{\rm c}$ (${H}_{\rm t}$), after 
making the rotating-wave approximation,
and the Liouville operators  $\mathcal{L_{\textrm{c}}}$ ($\mathcal{L_{\textrm{t}}}$)
are given in the basis $\{|0\rangle,|g\rangle,|1\rangle,|p\rangle,|r\rangle$\} as
\begin{widetext}

\begin{subequations}
\begin{eqnarray}
{H_{\rm (c/t)}} & = & \hbar\left(\begin{array}{ccccc}
-\omega_{10} & 0 & 0 & \Omega_{\rm R(c/t)}^{*}/2 & 0\\
0 & 0 & 0 & 0 & 0\\
0 & 0 & 0 & \Omega_{\rm R(c/t)}^{*}/2 & 0\\
\Omega_{\rm R(c/t)}/2 & 0 & \Omega_{\rm R(c/t)}/2 & \Delta_{p} & \Omega_{\rm B(c/t)}^{*}/2\\
0 & 0 & 0 & \Omega_{\rm B(c/t)}/2 & \Delta_{\rm r}\end{array}\right),\\
\label{eq:Hamiltonian}
\mathcal{L}&=&\left(\begin{array}{ccccc}
0.12\gamma_{p}\rho_{pp} & 0 & -\frac{\gamma_{01}}{2}\rho_{01} & -\frac{\gamma_{p}}{2}\rho_{0p} & -\frac{\gamma_{r}}{2}\rho_{0r}\\
0 & 0.56\gamma_{p}\rho_{pp} & 0 & -\frac{\gamma_{p}}{2}\rho_{gp} & -\frac{\gamma_{r}}{2}\rho_{gr}\\
-\frac{\gamma_{01}}{2}\rho_{10} & 0 & 0.32\gamma_{\rm p}\rho_{pp} & -\frac{\gamma_{p}}{2}\rho_{1p} & -\frac{\gamma_{1r}}{2}\rho_{1r}\\
-\frac{\gamma_{p}}{2}\rho_{p0} & -\frac{\gamma_{p}}{2}\rho_{pg} & -\frac{\gamma_{p}}{2}\rho_{p1} & -\gamma_{p}\rho_{pp}+\gamma_{r}\rho_{rr} & -\frac{\gamma_{p}+\gamma_{r}}{2}\rho_{pr}\\
-\frac{\gamma_{r}}{2}\rho_{r0} & -\frac{\gamma_{r}}{2}\rho_{rg} & -\frac{\gamma_{1r}}{2}\rho_{r1} & -\frac{\gamma_{p}+\gamma_{r}}{2}\rho_{rp} & -\gamma_{r}\rho_{rr}\end{array}\right).
\label{eq:Liouville}
\end{eqnarray}
\label{eq.HL1}
\end{subequations}
\end{widetext}

For Rb atoms we take into account the spontaneous emission from the intermediate level $|p\rangle$ to 
the ground level with a decay rate of $\gamma_{p}/2\pi=6.07~{\rm MHz}$  and
 the corresponding branching ratio of 0.56 to state $|g\rangle$, 0.32 to state $|1\rangle$ 
and 0.12 to state $|0\rangle$ as well as the decay from the Rydberg state $|r\rangle$ to 
the intermediate level $|p\rangle$ with rate $\gamma_{r}/2\pi=0.53~{\rm kHz}$.
 $\Delta_{p}$ is the intermediate level detuning, $\Delta_{{r}}=\Delta_{\rm AC}({\bf r})+\Delta_{\rm B}+\Delta_{\rm D}$
is the two photon detuning (see Sec. \ref{sec.technical}), $\omega_{10}/2\pi=6.83~{\rm GHz}$ is
the hyperfine ground state splitting; $\gamma_{1r}=\gamma_{r}+\gamma_{\rm ph}$
is the total dephasing of the Rydberg state relative to $|1\rangle$, 
$\gamma_{\rm ph}$ is the Rydberg dephasing rate due to magnetic field fluctuations and 
Doppler effects as shown in Eq. (\ref{eq:PhaseError});
and $\gamma_{01}$ is the dephasing of qubit states due to magnetic field fluctuations and atomic motion. 

Note that we do not include driving terms in Eq. (\ref{eq:Hamiltonian}) that couple the reservoir level $\ket{g}$ back to $\ket{p}$ and $\ket{r}$.
Doing so correctly would require adding additional Rydberg levels with different values of $m_j$ which would increase the numerical burden. Since any population in $\ket{g}$ is already fully counted as an error, including additional driving terms  would only reduce the final errors, and our results are reliably upper bounds on the gate error.

\begin{table}
\caption{Experimental parameters used in Ref. \cite{Zhang2010} as well as in the numerical simulations.
\label{tab:parameters} }
\begin{tabular}{|l|c|c|}
\hline 
Experimental parameter& symbol  & value\tabularnewline
\hline
FORT wavelength & $\lambda_{f}$ &  $1064~ {\rm nm}$ \tabularnewline
FORT waist & $w_{x,y}$ & $ 3.4 ~\mu{\rm m}$\tabularnewline
FORT trap depth & $U_{0}/k_{B}$ &  $4.5 ~\rm mK$\tabularnewline
FORT separation & $d$ & 8.7 $\mu\rm m$\tabularnewline
Atom temperature &$T_{\rm a}$ & $175 ~\mu\rm K$\tabularnewline
Rydberg level& $|r\rangle$ & $97d_{5/2}$\tabularnewline
Rydberg state radiative lifetime & $\tau$ & $320~\mu{\rm s}$\tabularnewline
Blockade shift  at $0~\mu{\rm K}$ & ${\sf B}_0/2\pi$ & $20~\rm MHz$\tabularnewline
Rydberg red power &&$2.4~\mu \rm W$\tabularnewline
Rydberg red waist &$w_{x/y,R}$ &$7.7~\mu{\rm m}$\tabularnewline
Rydberg red detuning &$\Delta_{p}/2\pi$& $-2~\rm GHz$\tabularnewline
Rydberg blue power && $13~\rm mW$\tabularnewline
Rydberg blue waist& $w_{x/y,B}$& $4.5~\mu{\rm m}$\tabularnewline
Rydberg red Rabi frequency & $\Omega_{\rm R}/2\pi$& $118 ~\rm MHz$\tabularnewline
Rydberg blue Rabi frequency& $\Omega_{\rm B}/2\pi$ & $39~\rm MHz$\tabularnewline
Rydberg Rabi frequency &$\Omega/2\pi$& $1.15~\rm MHz$\tabularnewline
Magnetic field fluctuation & $\sigma_{\rm B}$ & $2.5~\mu\rm T$\tabularnewline
Rydberg red power fluctuation && 1\%\tabularnewline
Rydberg blue power fluctuation && 2\%\tabularnewline
\hline
\end{tabular}
\end{table}

\subsection{Monte Carlo Simulations including technical errors}
\label{sec.masterexp}

In our numerical calculation, we consider two traps aligned along $z$ with separation
of $8.7~\mu{\rm m}$ as shown in Fig. \ref{fig:setup}a).
The atoms in each trap have temperature $T_{\rm a}$ and Gaussian spatial probability distribution
with variances of $\sigma_{x}$ , $\sigma_{y}$, and $\sigma_{z}$ as given in Sec. \ref{sec.technical}. 
Rydberg Rabi pulses are applied to the control or target atoms with Gaussian power fluctuations 
of FWHM of $1\%$ and $2\%$ for Rydberg red and blue lasers, respectively. 
An atom at position ${\bf r}=(x,y,z)$ with velocity ${\bf v}=(v_{x},v_{y},v_{z})$ experiences
a Rydberg excitation pulse with effective Rabi frequency and two-photon
detuning that depends on position and velocity, so $\Omega=\Omega({\bf r})$
and $\Delta_{r}=\Delta_{r}({\bf r},{\bf v})$ as described in Sec. \ref{sec.technical}. We also use a fit to the blockade curve of 
Fig. \ref{fig:blockadeshift} to account for variations in the two-atom interaction strength. 
Finally we can  monitor the effect of various error sources by switching them on and off and comparing the numerical results.

Numerical simulation of the CNOT gate demonstrated in Ref. \cite{Zhang2010} proceeds as follows.
We start with the initial density matrix for both atoms with population of $0.98$ in either $|1\rangle$
or $|0\rangle$ and population of $0.02$ in $|g\rangle$ to account for an optical pumping error of $1\%$ 
and atom loss of $1\%$ before the CNOT pulses. 
Then we perform Monte Carlo simulations of the experiment with the actual experimental parameters
from Ref. \cite{Zhang2010} as listed in  Table \ref{tab:parameters}.
Next, we solve the time evolution of the master equation (\ref{eq:master}) for the $C_Z$ pulse sequence of Fig. \ref{fig:sequence}a
with $\pi/2$ pulses on the target atom before and after the $C_Z$ to give a CNOT operation: 
 $\{(\frac{\pi}{2})_{\rm t},\pi_{\rm c},t_{\rm gap},(2\pi)_{\rm t},t_{\rm gap},\pi_{\rm c},(\frac{\pi}{2})_{\rm t}\}$ as in \cite{Zhang2010} 
and average over the four input states ($|00\rangle$, $|01\rangle$, $|10\rangle$ and $|11\rangle$) 
to get the averaged gate errors, where $(\frac{\pi}{2})_{\rm t}$ is a ground Raman $\pi/2$ pulse on the target atom, 
and $t_{\rm gap}=500~\rm ns$ is the switching time between control and target atom sites. 
Here, we treat the ground Raman pulse as an ideal unitary operation since it is substantially 
less sensitive to the various error sources than  operations involving Rydberg states.

\begin{table}
\caption{CNOT probability truth table and entanglement fidelity results. The upper section of the table gives  error budget values 
used in the numerical simulations. These values are based on laboratory measurements and theoretical estimates as described in 
\cite{Zhang2010}. The lower section gives measured results from \cite{Zhang2010}  and numerical values from solving Eq. (\ref{eq:master}).
\label{tab:results} }
\begin{tabular}{|lcc|}
\hline 
 & error & \tabularnewline
Parameter used in numerical simulation & budget&\tabularnewline
\hline
Optical pumping & 0.02&\tabularnewline
Atom loss before CNOT pulses & 0.02&\tabularnewline
Spontaneous emission & 0.018&\tabularnewline
Rydberg decay error & 0.003&\tabularnewline
Blockade error at $0~\mu{\rm K}$ & 0.0004&\tabularnewline
Blockade error at $175~\mu{\rm K}$ & 0.006&\tabularnewline
Doppler Broadening at $175~\mu{\rm K}$ & 0.003&\tabularnewline
Laser power fluctuation & 0.0001&\tabularnewline
Magnetic field fluctuation & 0.0002&\tabularnewline
\hline
  &measured&numerical \tabularnewline
&results&simulation\tabularnewline
\hline
Background loss (two atoms) & 0.19&\tabularnewline
CNOT trace loss ($1-{\rm Tr}[\rho_{\rm ct}]$) & 0.01&\tabularnewline
\hline
CNOT probability truth table&&\tabularnewline
\hline
raw fidelity&  0.74&0.75\tabularnewline
background loss corrected & 0.91&0.93\tabularnewline
background \& trace corrected & 0.92&0.93\tabularnewline
\hline
Bell state&& \tabularnewline
\hline
raw fidelity  & 0.58&0.54\tabularnewline
background loss corrected & 0.71&0.67\tabularnewline
background \& trace corrected & 0.71&0.67\tabularnewline
\hline
\end{tabular}
\end{table}

The probability truth table error is defined by $E=1-{\rm Tr}[\rho_{\rm ideal}^{T}\rho_{\rm ct}]$,
 where $\rho_{\rm ct}$ is the numerical simulation result. 
We should point out that the switching time $t_{\rm gap}$ is short enough that it has little effect on the CNOT truth table  fidelity, 
but has a strong effect on the entanglement fidelity because the motion of Rydberg excited atoms 
between excitation and deexcitation pulses leads to a stochastic phase that degrades the entanglement 
as was pointed out in \cite{Wilk2010}. 
Finally, we average over 100 evolutions of the master equation, 
and the final results are shown in Table \ref{tab:results}. 

To model the entanglement, we prepare the control atom in state 
$|c\rangle=\frac{1}{\sqrt{2}}(|0\rangle+i|1\rangle)$ and target atom in state $|t\rangle=|1\rangle$. 
Then we follow the same approach as for the CNOT truth table error analysis 
by solving the time evolution of the master equation (\ref{eq:master}). 
From the final density matrix after the pulse sequence, we can extract the Bell state  fidelity $F$  
defined as $F={\rm Tr}[\rho_{\rm ideal}^{T}\rho_{\rm ct}]$,
where $\rho_{\rm ideal}$ is associated with the maximally entangled Bell state
$|B_{1}\rangle=\frac{1}{\sqrt{2}}(|00\rangle+|11\rangle)$. 
We then  average over 50 evolutions of the master equation using Monte Carlo simulation
of all the error sources as mentioned before, and obtain the final Bell state fidelity $F=0.54$
without atom loss correction and fidelity $F=0.67$ after atom loss correction as shown in Table
\ref{tab:results}, which is consistent with our measured entangled fidelity of 0.71 in \cite{Zhang2010}. 
We assume that the atom loss due to collisions with untrapped background atoms 
is independent of the CNOT pulse sequence (about $4~\mu{\rm s}$) 
which is much shorter than the trap lifetime (several seconds), so the background loss is simply considered as
 a scaling factor for the final fidelity without atom loss.

\begin{figure}
\begin{centering}
\includegraphics[width=1\columnwidth]{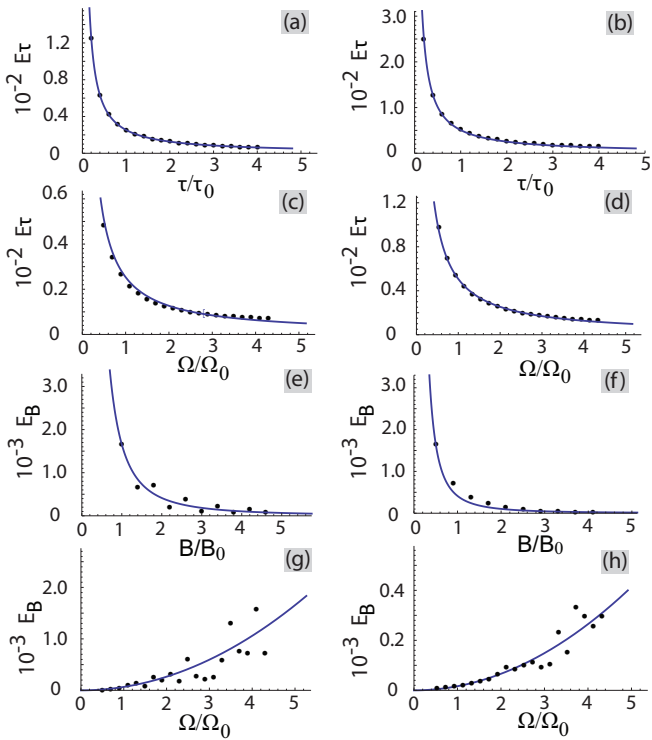}
\par\end{centering}
\caption{(color online) Comparison between intrinsic gate errors from Eq. (\ref{eq:IntrinsicError1}) 
for different parameters and numerical simulations.  
In the first column (a, c, e, g): ${\sf B}_{0}/2\pi=10~{\rm MHz}$, $\Omega_{0}/2\pi=1.15~{\rm MHz}$, 
$\tau_{0}=300~\mu s$; 
in the second column (b, d, f, h): ${\sf B}_{0}/2\pi=10~{\rm MHz}$, $\Omega_{0}/2\pi=0.575~{\rm MHz}$,
$\tau_{0}=300~\mu s$. 
$E_{\tau}$ is the decoherence error due to finite Rydberg lifetime (the first term in Eq. (\ref{eq:IntrinsicError1}))
and $E_{B}$ is the imperfect blockade error (the second term in Eq. (\ref{eq:IntrinsicError1})). 
The black dotted points are the results of the numerical simulations and the blue lines are the results from
Eq. (\ref{eq:IntrinsicError1}) with the same parameters.
 \label{fig:intrinsicerror}}
\end{figure}

In Figure \ref{fig:intrinsicerror}, we compare the numerical simulation results for the CNOT probability truth table 
with the analytical results of Eq. (\ref{eq:IntrinsicError1}) for different parameters.
 Both the decoherence error $E_{\tau}$ (the first term in Eq. (\ref{eq:IntrinsicError1}))
and imperfect blockade error $E_{B}$ (the second term in Eq. (\ref{eq:IntrinsicError1}))
agree well with the numerical results in the small gate error limit.
The total gate error with all the error sources is $6.5\%$ in agreement with the atom loss corrected fidelity of
 $F=0.92$ and the simple gate error analysis reported in \cite{Zhang2010}. 
As shown in Table \ref{tab:results}, the two main error sources limiting the gate fidelity are 
the spontaneous emission from state $|p\rangle$ and imperfect Rydberg excitation and 
blockade due to finite atomic temperature (not accounting for the atom loss before the CNOT pulses,
the imperfect optical pumping and other losses that are independent of the CNOT pulse sequence). 

The comparison of Monte Carlo master equation simulations with experimental data shows that the error sources listed in Table \ref{tab:results}
are able to account for measured  results with an accuracy of about 1\% as regards the CNOT truth table and about 5\% as regards the Bell state fidelity. The Bell state fidelity is much lower than that of the CNOT truth table  due to dephasing that occurs while the control atom is Rydberg excited. As has been discussed in \cite{Wilk2010,Saffman2011} the main sources of the Rydberg dephasing are magnetic field noise and Doppler effects. In order to significantly reduce these errors it will be necessary to work with colder atoms, less magnetic field noise, and faster Rydberg excitation pulses.

\subsection{Simulated Quantum Process Tomography}
\label{sec.masterlimit}

The experimental results obtained to date from Rydberg blockade experiments on pairs of atoms are far from predicted error thresholds for a practical fault-tolerant quantum computer which range from $10^{-4} - 10^{-2}$ in different models \cite{Knill2005,Raussendorf2007,*Aliferis2009,*Wang2011}. 
In order to characterize more completely the fidelity and usefulness of Rydberg blockade for quantum computing
applications  we need to perform Quantum Process Tomography (QPT) \cite{Chuang1997,*Poyatos1997,Nielsen2000}
of the Rydberg blockade mediated quantum blackbox process.
QPT has been demonstrated with several different physical systems 
including linear optics \cite{OBrien2004,*White2007},
trapped ions \cite{Riebe2006,*SXWang2010}, and superconducting circuits \cite{Yamamoto2010,*Bialczak2010}. 
Here, we perform numerical simulations of QPT with maximum likelihood estimation of 
 tomographically reconstructed density matrices  
\cite{OBrien2004,White2007} for the Rydberg-blockade $C_Z$ gate. 
We limit the simulations of intrinsic errors to the simpler $C_Z$ gate since it has been demonstrated\cite{Olmchenk2010,*Brown2011} that the additional single qubit pulses needed to implement a CNOT can be performed with errors at the $\sim 10^{-4}$ level.

Since our  goal is to determine the minimum possible gate error that can be reached using Rydberg blockade 
we only account for intrinsic gate errors as described in
Sec. \ref{sec.errors},  and assume all additional technical errors are negligible. This corresponds to a situation where the atoms are cooled to their motional ground state so there is no Doppler dephasing during Rydberg excitation,  position dependent variations in Rabi frequencies, or AC Stark shifts. We assume there is no spontaneous emission from the intermediate $|p\rangle$ level during Rydberg excitation. This could be achieved using one photon excitation of Rydberg $|p\rangle$ states, or by using sufficient laser power to detune very far from the intermediate $|p\rangle$ level.  We also assume that dephasing due to time varying magnetic fields is negligible.

Accounting only for intrinsic gate errors the analytical estimates of Sec. \ref{sec.errors}
show that $E\sim ({\sf B}\tau)^{-2/3}.$ At room temperature the Rydberg lifetime scales as $\tau\sim n^2$ with $n$ the principal quantum 
number and in
the heavy alkali atoms Rb and Cs  the van der Waals Blockade interaction scales as\cite{Saffman2008} ${\sf B}\sim n^{12}$. Thus we expect the 
gate error to scale as $E\sim n^{-28/3}$ so that choosing large  $n$ should give arbitrarily small errors.

\begin{figure}
\includegraphics[width=0.75\columnwidth]{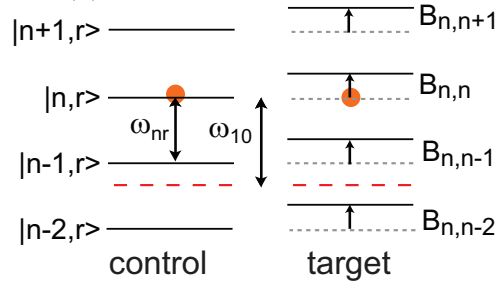}
\caption{(color online) Level diagram accounting for neighboring Rydberg levels of control and target atoms. The Rydberg laser is tuned to excite $|1\rangle$ to  $|n,r\rangle$. The red dashed 
lines show the energy that qubit state $|0\rangle$ is excited to,  and ${\sf B}_{n,n'}$ is the positive blockade shift between a control atom in $|nr\rangle$ and target atom in $|n'r\rangle$. 
\label{fig:optimumRydN}}
\end{figure}

This argument breaks down when $n \stackrel{>}{\sim} 100$ 
since  the energy spacing of levels $n$ and $n\pm 1$ becomes comparable to ${\sf B}$ or $\omega_{10},$ 
as shown in Fig. \ref{fig:optimumRydN}. This puts a limit on the effective blockade shift that can be achieved at large $n$ and limits the error floor.
 For the $C_Z$ pulse sequence acting on the four possible input states in the computational basis $\{|00\rangle,|01\rangle, |10\rangle, |11\rangle\}$ excitation is blocked three times due to $\omega_{10}$, once due to $\omega_{10}+{\sf B}$ and once due to ${\sf B}$ alone. 
It is necessary to choose the Rydberg level spacing and blockade shift such that the excitation suppression is as large as possible for all three cases. The above description is valid for Rydberg $p_{3/2}$ and $d_{5/2}$ states since by using $\sigma_+$ polarized light the qubit states are  only coupled 
to these Rydberg states \cite{gateerrornote1}.
For $s_{1/2}$ states the situation is worse since the Rydberg lasers simultaneously couple to both $s_{1/2}$ and $d_{3/2},d_{5/2}$ 
states.

In order to quantitatively account for coupling to more than one Rydberg level we have extended the basis used for simulations to the set  $\{|0\rangle,|g\rangle,|1\rangle,|r_2\rangle\,|r_1\rangle,|r\rangle\}$, where $|r_1\rangle, |r_2\rangle$ are additional Rydberg levels. Finding optimal states is now a multiparameter optimization problem. Details of how this is done, and the parameters of the chosen  $s_{1/2},p_{3/2}$ and $d_{5/2}$ states, are given in Appendix  \ref{app.1}. 
In this extended basis, but without the $|p\rangle$ level, the Hamiltonian and Liouville operators corresponding to 
Eqs.(\ref{eq.HL1}) are
\begin{widetext}
\begin{subequations}
\begin{eqnarray}
 H_{\rm (c/t)}  & = & \hbar \left(\begin{array}{cccccc}
-\omega_{10} & 0 & 0 & \Omega_{\rm (c/t)}^{''*}/2 & \Omega_{\rm (c/t)}^{'*}/2 & \Omega_{\rm (c/t)}^{*}/2 \\
0 & 0 & 0 & 0 & 0 & 0 \\
0 & 0 & 0 & \Omega_{\rm (c/t)}^{''*}/2 & \Omega_{\rm (c/t)}^{'*}/2 &  \Omega_{\rm (c/t)}^{*}/2 \\
\Omega_{\rm (c/t)}''/2 & 0 & \Omega_{\rm (c/t)}''/2 & -\omega_{r, r_2} & 0 & 0 \\
\Omega_{\rm (c/t)}'/2 & 0 & \Omega_{\rm (c/t)}'/2 & 0 & -\omega_{r,r_1} & 0 \\
\Omega_{\rm (c/t)}/2 & 0 & \Omega_{\rm (c/t)}/2 & 0 & 0 & 0 \end{array}\right),
\label{eq:Hamiltonian2}\\
\mathcal{L}_{\rm (c/t)}&=&\gamma_r\left(\begin{array}{cccccc}
\frac{1}{8}\rho_{rT} & 0 & 0  & -\frac{1}{2} \rho_{0r_2} & -\frac{1}{2}\rho_{0r_1}  & -\frac{1}{2} \rho_{0r} \\
0 & \frac{3}{4}\rho_{rT} & 0 & -\frac{1}{2} \rho_{gr_2}  & -\frac{1}{2} \rho_{gr_1} & -\frac{1}{2} \rho_{gr} \\
0  & 0 & \frac{1}{8}\rho_{rT} & -\frac{1}{2} \rho_{1r_2}  & -\frac{1}{2} \rho_{1r_1} & -\frac{1}{2} \rho_{1r} \\
-\frac{1}{2} \rho_{r_2 0}  & -\frac{1}{2} \rho_{r_2 g}  & -\frac{1}{2} \rho_{r_2 1}  & -\rho_{r_2r_2}  & -\rho_{r_2 r_1} &  -\rho_{r_2r}\\
-\frac{1}{2} \rho_{r_1 0}  & -\frac{1}{2} \rho_{r_1 g}  & -\frac{1}{2} \rho_{r_1 1}  & -\rho_{r_1r_2}  &
 -\rho_{r_1 r_1} &  -\rho_{r_1 r}\\
-\frac{1}{2} \rho_{r 0}  & -\frac{1}{2} \rho_{r g}  & -\frac{1}{2} \rho_{r 1}  & -\rho_{rr_2}  & -\rho_{r, r_1} &  -\rho_{r r} 
\end{array}\right).
\label{eq:Liouville2}
\end{eqnarray}
\label{eq:HL2}
\end{subequations}
\end{widetext}
with $\rho_{rT}=\rho_{rr}+\rho_{r_1r_1}+\rho_{r_2r_2}$ the total Rydberg excited population. 
The two-atom operators are 
\begin{equation}
{H}_{\rm ct}={H}_{\rm c}\otimes I_{\rm t}+I_{\rm c}\otimes{H}_{\rm t}+{\sf B}_{\rm ct},
\label{eq.Hct2}
\end{equation}
$\mathcal{L}_{\rm ct}=\mathcal{L}_{\rm c}\otimes I_{\rm t}+I_{\rm c}\otimes\mathcal{L}_{\rm t}$, 
$I_{\rm (c/t)}$ are $6\times6$ identity matrices, and 
${\sf B}_{\rm ct} = \hbar\, {\rm diag}[0_{21}, {\sf B}_{\rm r_2r_2}, {\sf B}_{\rm r_2r_1}, {\sf B}_{\rm r_2 r}, 0, 0, 0,$ $ {\sf B}_{\rm r_1 r_2}, {\sf B}_{\rm r_1 r_1}, {\sf B}_{\rm r_1 r}, 0, 0, 0, {\sf B}_{\rm r r_2}, {\sf B}_{\rm r r_1}, {\sf B}_{\rm r r}] $ is a Rydberg blockade matrix where 0$_{21}$ is a 21 zero list.
We assume that the Rydberg states decay directly back to the 8 ground sublevels of Rb with equal branching ratios of $1/8$.  
For Cs atoms the factors of $1/8,3/4,1/8$ on the diagonal of (\ref{eq:Liouville2}) become $1/16,7/8,1/16$. The Rabi frequencies for Rydberg excitations to states $|r\rangle, |r_1\rangle, |r_2\rangle$ are taken to be equal $\Omega=\Omega'=\Omega''$ for Rydberg $np$ and $nd$ cases. For Rydberg $ns$ states we have $|r\rangle=|ns\rangle$, $|r_1\rangle=|n-1,s\rangle$, $|r_2\rangle=|n-2,d\rangle$ for which 
$\Omega'\simeq\Omega, \Omega''=1.31 \Omega$. Further details are given in Appendix \ref{app.1}. For simplicity we assume the decay rate $\gamma_r$ is the same for all Rydberg levels. It is straightforward to include state dependent decay rates in the code, but this has a negligible impact on the results since there is very small excitation of the secondary Rydberg states. For simplicity we use the decay rate of the targeted Rydberg state for all states.

As discussed in Ref. \cite{Gilchrist2005}, there is no universally agreed upon measure for comparing 
real and idealized quantum processes\cite{Nielsen2002,*Pedersen2007}. A widely used measure of quantum process fidelity is the trace overlap fidelity $F_{\rm O}$, or error $E_{\rm O}=1-F_{\rm O}$ 
which are based on the trace overlap between  ideal and experimental (in our case simulated) $\chi$ process matrices. 
Another error measure $E_{\rm D}$ is defined as the trace distance between
the ideal and simulated  matrices. 
We have quantified  process errors using the trace overlap and trace distance  as
\begin{subequations}
\begin{eqnarray}
E_{\rm O} & = &1- {\rm Tr}^2\left[\sqrt{\sqrt{\chi_{\rm sim}}\chi_{\rm id}\sqrt{\chi_{\rm sim}}}\right],
\label{eq:error2}\\
E_{\rm D} & = & \frac{1}{2}{\rm Tr}\left[\sqrt{\left(\chi_{\rm id}-\chi_{\rm sim}\right)^{\dagger}\left(\chi_{\rm id}-\chi_{\rm sim}\right)}\right],
\label{eq:distance}
\end{eqnarray}
\label{eq:errormeasure}
\end{subequations}
where $\chi_{\rm id}$ is the ideal process matrix and 
$\chi_{\rm sim}$ is the simulated physical $\chi$-matrix found from QPT accounting for  intrinsic gate errors as described by Eqs. (\ref{eq:HL2},\ref{eq.Hct2}). We use a maximum likelihood estimator to extract a physical $\chi$ matrix from the QPT simulations\cite{OBrien2004,*White2007}.

\begin{table*}[!t]
\caption{Gate errors from simulated QPT for several Rydberg states of 
$^{87}$Rb and $^{133}$Cs. The reported errors are $E_{\rm cb}$,  the analytical estimate found in Appendix \ref{app.1} using computational basis states, trace loss, which is the sum of populations outside the computational basis at the end of the gate sequence, $E_{\rm O}(E_{\rm D})$ trace overlap(distance) errors from Eqs. (\ref{eq:errormeasure}) using the Jaksch et al. pulse sequence (Fig. \ref{fig:sequence}a),  and $E_{\rm D}'$ the trace distance error found using the modified pulse scheme of Fig. \ref{fig:sequence}b. For the $s$ and $p$ states we used the optimal Rabi frequencies found in Appendix A. For the $d$ states we
found that the $E_{\rm O}, E_{\rm D}'$ errors were reduced by about 25\% by using Rabi frequencies about 15\% lower than those estimated in the Appendix.   
}
\begin{tabular}{|l|llll|llll|}
\hline
& $^{87}$Rb & &&&Cs&&&\\
\hline 
Rydberg state &$76p_{3/2}$ &$124p_{3/2}$&$123d_{5/2}$&$82s_{1/2}$&  $70p_{3/2}$& $112p_{3/2}$ &$112d_{5/2}$ &$80s_{1/2}$\tabularnewline
Rabi frequency $\Omega/2\pi~\rm (MHz)$& 38.5 &16.3&15.3&19.2&47.1&19.5&20.4&21.4 \tabularnewline
\hline
 $E_{{\rm  cb}}$   & $.00015$ &$.00013$&$.00016$&$.00032$& $.00013$&.00011&.00018&.00032 \tabularnewline
 trace loss   & $.00050$ &$.00062$&$.0010$&$.0013$& $.00046$&$.00066$&$.00073$&$.0013$ \tabularnewline
 $E_{{\rm  O}}$   & $.0012$ & $.0013$&$.0018$& $.0023$&.0011&.0015&.0014&.0025  \tabularnewline
 $E_{\rm D}$  & $.0058$ &.0047&.0041&.0071&.0032 &.0050&.0067&.0081 \tabularnewline
 $E_{\rm D}'$  & $.0012$ &.0015&.0016&.0024&.0013&.0020 &.0018 &.0028\tabularnewline
\hline
\end{tabular}
\label{tab:QPTerrors}
\end{table*}

In Table \ref{tab:QPTerrors} we present the errors found from simulated QPT for the  atomic states in Table \ref{tab.Rydbergsdp}.
The process tomography errors tend to be 5-10 times larger than $E_{\rm cb}$ which are the errors estimated in Appendix \ref{app.1} for 
two-qubit product states in the  computational basis. This is to be expected since the analytical estimates are derived from  
the probabilities of the gate succeeding, and do not account for output state phase errors.  The trace loss quantifies the population in states outside the computational basis at the end of the gate sequence. These errors are due to spontaneous emission from Rydberg states and imperfect blockade which leaves atoms Rydberg excited at the end of the gate. Trace loss errors account for about half of the process error. While the process error based on trace overlap $E_{\rm O}$ is less than $0.003$ for all states listed, the error as measured by the trace distance $E_{\rm D}$ is significantly larger.

\begin{figure}[!t]
\begin{centering}
\includegraphics[width=1\columnwidth]{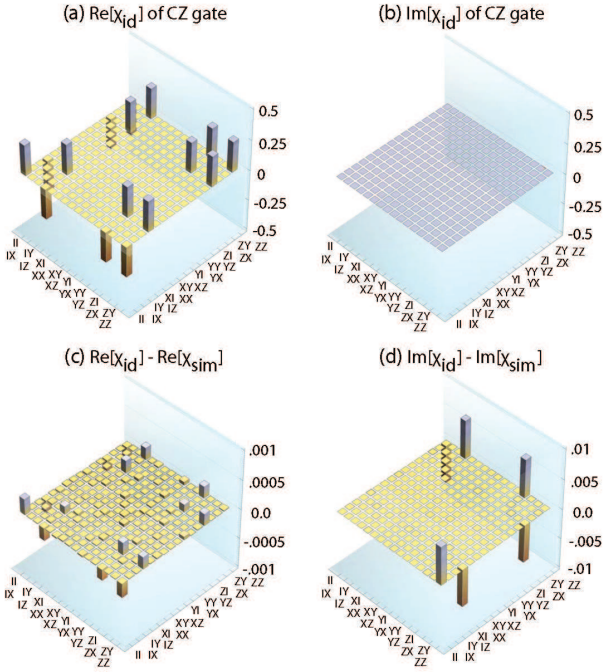}
\par\end{centering}
\caption{(color online) Ideal $\chi_{\rm id}$ matrix and the difference between ideal $\chi_{\rm id}$
and simulated $\chi_{\rm sim}$ matrices for QPT of the C$_Z$ gate using the Jaksch, et al. pulse sequence\cite{Jaksch2000}  for $^{87}$Rb $112p_{1/2}$ states 
with $\Omega/2\pi=27.6~{\rm MHz}$. 
(a) Real part of  $\chi_{\rm id}$, 
(b) Imaginary part of $\chi_{\rm id}$, 
(c) difference of real parts, ${\rm Re}[\chi_{\rm id} - \chi_{\rm sim}]$, and 
(d) difference of imaginary parts ${\rm Im}[\chi_{\rm id} - \chi_{\rm sim}]$.
\label{fig:chimatrix} }
\end{figure}

Some insight can be gleaned into why the trace distance gives larger errors than the trace overlap  as follows.  The Jaksch et al.  pulse sequence produces an imperfect $C_Z$ gate which can be written in the computational basis as 
\begin{eqnarray}
C_{Z} & = & \left(\begin{array}{cccc}
1 & 0 & 0 & 0\\
0 & -1 & 0 & 0\\
0 & 0 & -1 & 0\\
0 & 0 & 0 & -e^{i\phi}\end{array}\right),
\label{eq:CzIm}
\end{eqnarray}
where $\phi=\pi\Omega/2{\sf B}$ is a small phase error in the  strong blockade limit. 
As an example Fig. \ref{fig:chimatrix} shows the ideal $\chi$-matrix and the difference 
between the ideal and simulated $\chi$-matrices for the standard pulse sequence leading to (\ref{eq:CzIm}).
From Fig. \ref{fig:chimatrix}(c) and (d), we can see that the error in the 
imaginary part of the 
 $\chi$-matrix is much larger than that in the real part of the $\chi$-matrix. 
This is due to the fact that the real part is related to the amplitude errors 
for Rydberg blockade
which are proportional to $\phi^{2}$, but the imaginary
part is related to the phase error which scales as $\phi$.

Following the procedure for QPT in Ref. \cite{Nielsen2000} we can calculate the ideal and 
imperfect $\chi$-matrices from (\ref{eq:CzIm}), and using Eqs.  (\ref{eq:errormeasure}) we find the trace overlap and trace distance errors  
\begin{subequations}
\begin{eqnarray}
E_{\rm O} & = & \frac{3}{8}\left[1-\cos(\phi)\right]\backsimeq\frac{3}{16}\phi^{2}\sim\left(\frac{\Omega}{\sf B}\right)^{2},
\label{eq:errorTh1}
\\
E_{\rm D} & = & \frac{\sqrt{3}}{2}\sin(\phi/2)\backsimeq\frac{\sqrt{3}}{4}\phi\sim\frac{\Omega}{\sf B}.
\label{eq:distanceTh}
\end{eqnarray}
\end{subequations}
We see that $E_{\rm O}\sim \phi^2$ while $E_{\rm D}\sim \phi$ which verifies that the trace overlap is not sensitive to the imaginary part of $\chi$, which has linear phase errors, whereas the trace distance is sensitive to these errors.

Using the modified sequence of Fig. \ref{fig:sequence}b we can correct the leading order linear term in the phase error. Doing so has negligible effect on the trace overlap since it is only sensitive to amplitude errors at ${\mathcal O}(\phi^2)$. We do not report trace overlap errors for the modified pulse sequence in Table \ref{tab:QPTerrors} since they are unchanged. However there is a large reduction in the trace distance error using the modified pulse sequence as can be seen from the values of $E_{\rm D}'$ in the last row of the table. 
It has been common practice in experimental studies of quantum gate process fidelity to use the trace overlap as a reliable measure of the gate fidelity. The results shown in Table \ref{tab:QPTerrors} highlight the fact that  the trace overlap may give an overly optimistic view of the gate performance, since the trace distance gives larger errors. Identifying what type of errors are present and finding ways to minimize them, as we have done here
using a modified pulse sequence, is facilitated by checking several error measures.

Finally we note that besides the intrinsic gate error sources, the dipole-dipole interaction will cause a momentum kick
 to both atoms \cite{Jaksch2000} which can excite a trap state without changing 
the internal state of the atoms when they are in Rydberg states. 
The perturbative transition probability is bounded by $p_{k}<(3\eta{\Omega}^{2}\Delta t/8{\sf B})^{2}/2$
with $\Delta t=2\pi/\Omega$ and $\eta=a/d\ll1,$ where $a$ is the initial width of the atomic wave function 
determined by the trap and $d$ is the trap separation. 
For typical gate parameters  and $\eta=1/50$ we find that $p_{k}<3\times 10^{-7}$ 
for Rb Rydberg $124p_{3/2}$ states which is much smaller than the QPT errors in Table \ref{tab:QPTerrors}. 
Thus errors due to momentum transfer between Rydberg excited atoms have a negligible effect on the gate fidelity.

\section{Discussion}
\label{sec.discussion}

An important motivation for performing detailed calculations of  gate errors is to determine if Rydberg blockade could be used to build a fault tolerant, large scale quantum computing device. In order to answer that question it is necessary to make a connection between the fidelity measures, and the error limits for fault tolerant architectures that have been calculated theoretically. Threshold calculations typically proceed 
by assuming that an ideal unitary operator $U_{\rm id}$ describing the time evolution of the quantum circuit is corrupted by an error operator $U_{\rm er}$ with a small probability $\epsilon$, $0\le\epsilon\le 1$ so that the actual gate is described probabilistically as 
$$
U=(1-\epsilon)U_{\rm id}+ \epsilon U_{\rm er}.
$$ 
Depending on what assumptions are made about the types of error operators $U_{\rm er}$ that may occur, possible correlations between 
errors at different sites, and the overall system architecture, different threshold values $\epsilon_{\rm th}$ can be found. Provided 
$\epsilon<\epsilon_{\rm th}$ it is in principle feasible to build an arbitrarily large quantum processor. Calculations that make a minimum number of assumptions about  $U_{\rm er}$ result in very low thresholds, 
$\epsilon_{\rm th}\sim 10^{-5}$\cite{Aliferis2006}. Other calculations  that make more restrictive assumptions   result in higher thresholds.   For example  $\epsilon_{\rm th}> 0.03$ in a model where the $U_{\rm er}$ are Pauli operators\cite{Knill2005}. 

In order to relate the fidelity measures  to thresholds for fault tolerance it is necessary to make explicit the connection between the process fidelity and the error probability $\epsilon$, as has been done for photonic quantum gates\cite{Weinhold2008}.  We can  estimate  the lower bound on $\epsilon$ given a process fidelity as follows. Replace the process matrix in Eqs. (\ref{eq:errormeasure}) by $\chi_{\rm sim}\rightarrow (1-\epsilon)\chi_{\rm id}+\epsilon \chi_{\rm er}$ where $\chi_{\rm er}$ is the process matrix corresponding to the operator $U_{\rm er}$. 

To derive lower bounds on $\epsilon$ we substitute the modified expression for $\chi_{\rm sim}$ into (\ref{eq:errormeasure}a) to get 
\begin{eqnarray}
E_{\rm O} &=& \epsilon{\rm Tr}^2\left[\sqrt{\sqrt{(\chi_{\rm id}-\chi_{\rm er})}\chi_{\rm id}\sqrt{(\chi_{\rm id}-\chi_{\rm er})}}\right]+O(\epsilon^2)\nonumber.
\end{eqnarray}
The right hand side is maximized for  $\chi_{\rm er} =-\chi_{\rm id}$ and assuming small $\epsilon$ we find $E_{\rm O}\le 2\epsilon.$
We can therefore use the trace overlap to bound the error probability from below according to

\begin{equation}
\epsilon_{\rm O}  \ge  \frac{E_{\rm O}}{2}.
\end{equation}
Following the same steps for the trace distance gives 
\begin{equation}
\epsilon_{\rm D}  \ge  E_{\rm D}.
\end{equation}
 These bounds result from assuming $\epsilon\ll 1$ and a worst case error process with ${\rm Tr}[\chi_{\rm id}\chi_{\rm er}]=-1$. Our calculated fidelity errors given in Table \ref{tab:QPTerrors} are ${\mathcal O}(10^{-3})$ and we therefore have placed lower bounds on $\epsilon$ of ${\mathcal O}(10^{-3})$ which is below the threshold for some fault tolerant architectures. 

Unfortunately this does not prove fault tolerance.
We have bounded $\epsilon$ from below, but the actual $\epsilon$ for our gates may be higher. In addition, 
 threshold calculations make assumptions about the types of errors that may occur, whereas our calculations of process fidelities are based on an independent physical model of the gate.  In order to claim fault tolerance we would have to verify that the errors occurring in our simulations are compatible with the assumptions made in the threshold calculations. This has only
been attempted  for linear optics quantum gates\cite{Weinhold2008} and is beyond the scope of the present paper.   All we can say based on the results obtained here is that it is plausible that the fidelity of Rydberg blockade gates is sufficient to meet the threshold for fault tolerance in an appropriate architecture, but this has not been explicitly demonstrated. 

In conclusion we have performed a detailed analysis and numerical simulation of our recent demonstration
of a Rydberg blockade mediated CNOT gate between two individually addressed neutral atoms. 
Good agreement between the model and experimental results allows us to identify
the leading error sources limiting the CNOT truth table  fidelity as imperfect state preparation, spontaneous emission 
from the intermediate state during two-photon Rydberg excitation and imperfect Rydberg excitation 
and blockade due to variations of the atomic position  at finite  temperature. 
The fidelity of entangled Bell states created so far with  Rydberg blockade  is predominantly 
limited by ground-Rydberg dephasing due to Doppler broadening and magnetic field noise.
 
We have also found intrinsic error limits for Rydberg states which are accessible by one or two photon excitation through dipole allowed transitions. 
 We show that the gate error 
cannot be made arbitrarily small by addressing higher lying Rydberg levels due to off-resonant coupling to neighboring levels which reduces the blockade effect. We identified the optimum blockade strength 
in the presence of neighboring Rydberg levels and showed using simulated QPT that for both $^{87}$Rb and Cs atoms states can be found with process errors below $0.002$, provided we use a modified pulse sequence to correct small phase errors. The phase error correction assumes that the atoms are in the motional ground state of the optical traps. 
Our identification of optimum parameters including coupling to neighboring Rydberg levels is only approximate and  
it may be possible to further reduce the gate error with a more extensive parameter search.

While we have focused on the Rydberg blockade mechanism, the direct-interaction Rydberg phase gate\cite{Jaksch2000}, which uses simultaneous excitation of both atoms to a Rydberg level, may also be a route to high fidelity operation. Recent analysis of this gate using optimal control 
theory has identified parameters for which the gate error approaches $10^{-3}$ although 
a rigorous process error was not calculated\cite{Muller2011}. We note that also the phase gate which operates with the ordering $\Omega\gg {\sf B}$
will be subject to a limit on how high $n$ can be due to off-resonant excitation of  neighboring Rydberg levels, as illustrated for the blockade gate in Fig. \ref{fig:optimumRydN}.

Our error results assume operation in a room temperature environment. The lifetimes of the $p_{3/2}$ states  increase by about a factor of four in a 4K He cryostat\cite{Beterov2009,*Beterov2009b}, which would result in 
a reduction of the gate error by a factor of $\sim 2.5$ to a level below  $ 10^{-3}$. Even lower error levels could in principle be reached using circular Rydberg states that have orders of magnitude longer lifetimes, although there are serious technical challenges connected with high fidelity excitation and de-excitation of these states.

\begin{acknowledgments}
M.S. would like to thank Andrew White for helpful discussions. 
This work was supported by NSF award PHY-1005550, the IARPA MQCO program through ARO contract W911NF-10-1-0347, 
and DARPA.
\end{acknowledgments}


%


\appendix

\section{CNOT truth table error estimates with multiple Rydberg levels}

\label{app.1}

In this appendix we present analytical estimates for the CNOT truth table including  off-resonant excitation of multiple Rydberg levels. These estimates were used to find  parameters for the process tomography calculations in Sec. \ref{sec.master}.

 \subsection{Rydberg $np_{3/2}$ states}
\label{sec.Rydbergp}

\begin{figure}[!h]
\begin{centering}
\includegraphics[width=1\columnwidth]{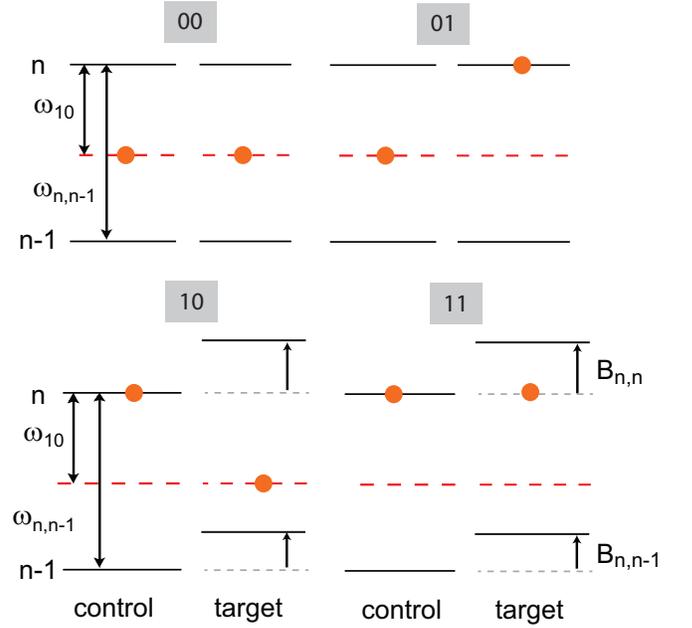}
\par\end{centering}
\caption{(color online) Off-resonant Rydberg excitation for $p$ states with $\omega_{n,n-1}>\omega_{10}$. The filled circles indicate the energies where the atoms are excited to with the long-dashed red lines corresponding to excitation of an atom starting in $\ket{0}$.
\label{fig:p_rydlevels} }
\end{figure}

Alkali atom $np_{1/2}$ or $np_{3/2}$ levels can be reached by one-photon excitation from the ground state. The fine structure splitting of high $n$ Rydberg levels is relatively small, only 94 MHz for the Rb $100p$ states. This small splitting is problematic since resonant coupling to say $100p_{1/2}$ with a Rabi frequency of $\sim 10~\rm MHz$ would give errors at the $0.01$ level due to off-resonant coupling to $100p_{3/2}$. We therefore assume that the qubit state $\ket{1}$ is mapped onto the stretched ground state $f_+=I+1/2$, $m_f=f$ before and after Rydberg operations. When the stretched state is excited with $\sigma_+$ light angular momentum selection rules prevent coupling to $np_{1/2}$. We therefore only account for coupling to 
 $np_{3/2}, m_j=3/2$ states.

Referring to Fig. \ref{fig:p_rydlevels}  we assume the spacing between neighboring levels $\omega_{n,n-1},$ satisfies 
$\omega_{n,n-1}>\omega_{10}$. This corresponds to Rydberg levels with $n\stackrel{<}{\sim}100$ for the heavy alkalis. The leading contributions to blockade errors for the computational basis states are
\begin{eqnarray}
E_{00} &=& \frac{3}{2}\left[\frac{\Omega^2}{\omega_{10}^2}+ \frac{\Omega^2}{(\omega_{n,n-1}-\omega_{10})^2}\right] \nonumber\\
E_{01} &=&  \frac{2}{3}E_{00}\nonumber\\
E_{10} &=&\frac{1}{2}\left[ \frac{\Omega^2}{(\omega_{n,n-1}-{\sf B}_{n,n-1}-\omega_{10})^2} + \frac{\Omega^2}{(\omega_{10}+{\sf B}_{n,n})^2}\right]\nonumber\\
E_{11} &=& \frac{1}{2}\left[\frac{\Omega^2}{{\sf B}_{n,n}^2} + \frac{\Omega^2}{(\omega_{n,n-1}-{\sf B}_{n,n-1})^2}\right].\nonumber
\end{eqnarray}
The average blockade error is $E_B=\frac{1}{4}(E_{00}+E_{01}+E_{10}+E_{11}).$ We introduce two dimensionless parameters $a=\omega_{n,n-1}/\omega_{10}$ and $b={\sf B}_{n,n}/\omega_{10}.$ Parameter $a$ takes on discrete values as a function of Rydberg level $n$ while  $b$ can be adjusted to minimize the error at fixed $n$ by changing the interatomic separation $d$. The blockade interaction between levels of different $n$ is also a function of $d$, but to a good approximation we can put ${\sf B}_{n,n-1}=b' {\sf B}_{n,n}$, with $b'$ a constant independent of $d$. We neglect contributions from coupling to level $n+1$ since the solutions found below have $b\sim 0.5$ and the average error from $n+1$ states only contributes at the 10\% level. 

\begin{figure}[!t]
\begin{centering}
\includegraphics[width=1\columnwidth]{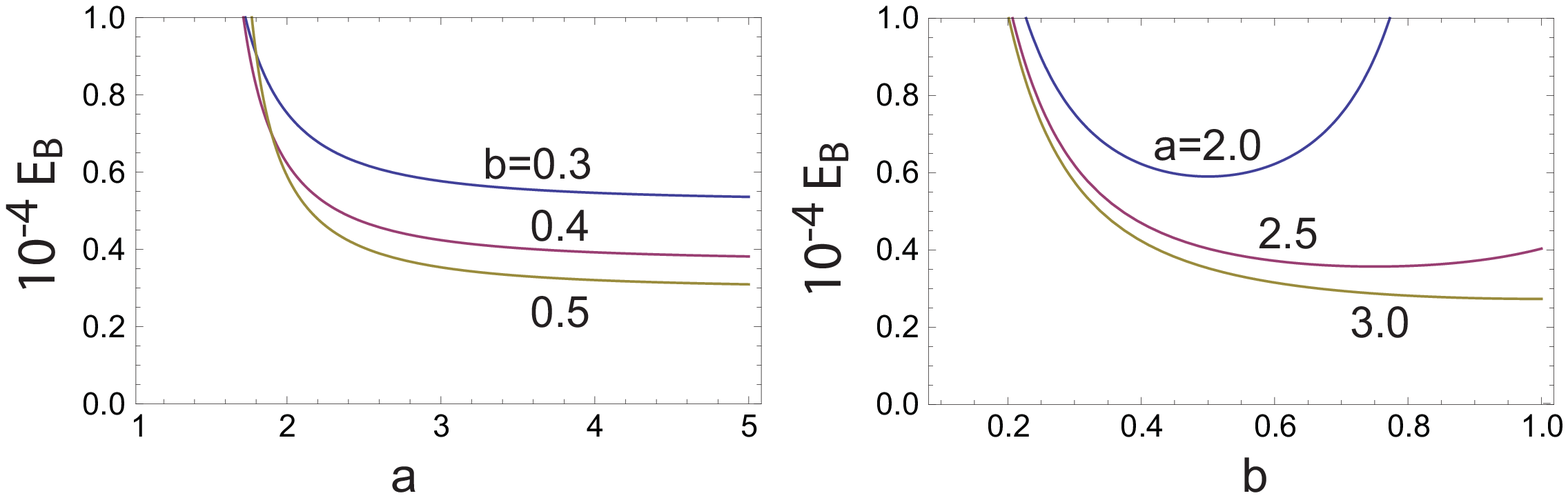}
\par\end{centering}
\caption{(color online) Blockade error scaling for 
$p$ states with $\omega_{n,n-1}>\omega_{10}$ as a function of $a,b$  with $\Omega/\omega_{10}=0.005$
 and $b'=1$.
\label{fig.perror1} }
\end{figure}

The blockade error is minimized by ensuring that all undesired excitations are detuned as much as possible. This corresponds to $\omega_{n,n-1}$ very large but this is not a useful solution since it implies small $n$ and large spontaneous emission errors. In order to find a reasonable value for $n$ consider Fig. \ref{fig.perror1} which shows the blockade error for selected values of $a,b$. We see that for $a\sim 2.5$ and $b\sim 0.5$ the error is not far from the minimum possible. The scaled blockade shift $b$ could be made larger, but this would require very small values of the separation $d$ which implies difficulty in individual addressing of the atoms. 

The above conditions are matched quite closely for $^{87}$Rb (Cs) using $np_{3/2}$ states with $n=76(70).$ 
For $^{87}$Rb with $n=76$ at $d=1.8~\mu\rm m$ we have 
 $\omega_{76,75}=2\pi\times 17.0~\rm GHz$, $a=2.49$,  $b=0.51$, $b'=1.07$ which gives $E_B=0.63\times 10^{-4}$ at $\Omega/\omega_{10}=0.005$.
For Cs with $n=70$ at $d=1.35~\mu\rm m$ we have 
 $\omega_{70,69}=2\pi\times 23.0~\rm GHz$, $a=2.50$,  $b=0.48$, $b'=1.23$ which gives $E_B=0.66\times 10^{-4}$ at $\Omega/\omega_{10}=0.005$.

We can estimate the CNOT truth table error averaged over the computational basis states using the same procedure as in Sec. \ref{sec.errors}.
Including the spontaneous emission errors from Eq. (\ref{eq:IntrinsicError1}), neglecting corrections of order $\Omega^2/\omega_{10}^2, \Omega^2/{\sf B}^2,$ the average error is
$$
E=\frac{7\pi}{4\Omega\tau} + E_{B0}\Omega^2
$$
where we have written $E_B=E_{B0}\Omega^2$. The optimum Rabi frequency is $\Omega_{\rm opt}=\left(\frac{7\pi}{8} \right)^{1/3}\frac{1}{(E_{B0}\tau)^{1/3}}$ which gives the minimum  error for the computational basis states
\begin{equation}
E_{\rm cb}=\frac{3\left(7\pi\right)^{2/3}}{4}\frac{E_{B0}^{1/3}}{\tau^{2/3}}.
\label{eq.Eminpryd}
\end{equation}
For $^{87}$Rb $76p_{3/2}$ with $\tau=223~\mu\rm s$ we find $\Omega_{\rm opt} = 2\pi\times 38.5~\rm MHz$ ($\Omega_{\rm opt}/\omega_{10}=0.006$) and $E_{\rm cb}=1.5\times 10^{-4}$.
For Cs $70p_{3/2}$ with $\tau=211~\mu\rm s$ we find $\Omega_{\rm opt} = 2\pi\times 47.1~\rm MHz$ ($\Omega_{\rm opt}/\omega_{10}=0.005$) and $E_{\rm cb}=1.3\times 10^{-4}$.


\begin{figure}[!t]
\begin{centering}
\includegraphics[width=1\columnwidth]{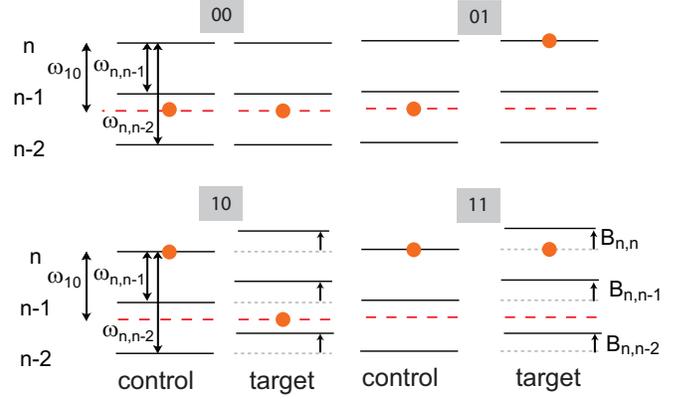}
\par\end{centering}
\caption{(color online) (color online) Off-resonant Rydberg excitation for $p$ states with $\omega_{n,n-1}>\omega_{10}>\omega_{n,n-2}$. The filled circles indicate the energies where the atoms are excited to with the long-dashed red lines corresponding to excitation of an atom starting in $\ket{0}$.
\label{fig:p_rydlevels2} }
\end{figure}

It is also possible to consider states with higher $n$ such that
$\omega_{n,n-2}>\omega_{10}>\omega_{n,n-1}$. In this case, which is shown  in Fig. \ref{fig:p_rydlevels2}, we must include off-resonant coupling to  a third Rydberg level $n-2$. The effective blockade shift is now smaller than for the lower $n$ states but there is the advantage that the Rydberg lifetime is longer. The blockade errors for the computational basis states are 

\begin{figure}[!t]
\begin{centering}
\includegraphics[width=.9\columnwidth]{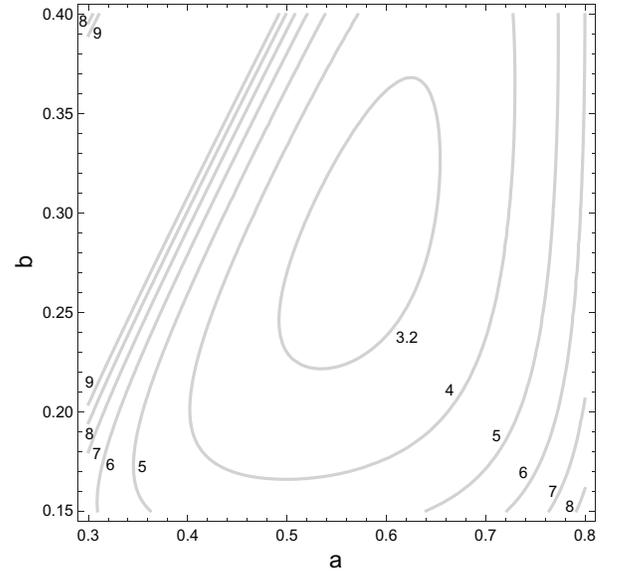}
\par\end{centering}
\caption{(color online) Blockade error scaling for $p$ states with $\omega_{n,n-2}>\omega_{10}>\omega_{n,n-1}$ as a function of $a,b$  with $\Omega/\omega_{10}=0.005$, $\omega_{n,n-2}=2\omega_{n,n-1}$,  and  $b'=b''=1$ (${\sf B}_{n,n-2}=b'' {\sf B}_{n,n}$).
The contours are labeled with $E_B$ in units of $10^{-4}$.
\label{fig:p_ryderror2} }
\end{figure}

\begin{eqnarray}
E_{00} &=& \frac{3}{2}\left[\frac{\Omega^2}{(\omega_{10}-\omega_{n,n-1})^2}+ \frac{\Omega^2}{(\omega_{n,n-2}-\omega_{10})^2}\right] \nonumber\\
E_{01} &=& \frac{2}{3}E_{00}  \nonumber\\
E_{10} &=&\frac{1}{2}\left[ \frac{\Omega^2}{(\omega_{10}-\omega_{n,n-1}+{\sf B}_{n,n-1})^2}\right.\nonumber\\
&& \left. + \frac{\Omega^2}{(\omega_{n,n-2}-\omega_{10}-{\sf B}_{n,n-2})^2}\right]\nonumber\\
E_{11} &=& \frac{1}{2}\left[\frac{\Omega^2}{{\sf B}_{n,n}^2} + \frac{\Omega^2}{(\omega_{n,n-1}-{\sf B}_{n,n-1})^2}\right].\nonumber
\end{eqnarray}
The average blockade error $E_B=\frac{1}{4}(E_{00}+E_{01}+E_{10}+E_{11})$ is shown in Fig. \ref{fig:p_ryderror2}. 

The error is minimized for $a\sim 0.57$ and $b\sim0.28$. These conditions are matched quite closely for $^{87}$Rb (Cs) using $np_{3/2}$ states with $n=124(112).$
For $^{87}$Rb with $n=124$ at $d=4.5~\mu\rm m$ we have 
 $\omega_{124,123}=2\pi\times 3.73~\rm GHz$, $\omega_{124,122}=2\pi\times 7.55~\rm GHz$, 
$a=0.55$,  $b=0.29$, $b'=1.05$, $b''=1.06$ which gives $E_B=3.2\times 10^{-4}$ at $\Omega/\omega_{10}=0.005$.
For Cs with $n=112$ at $d=3.2~\mu\rm m$ we have 
 $\omega_{112,111}=2\pi\times 5.23~\rm GHz$, $\omega_{112,110}=2\pi\times 10.6~\rm GHz$, 
 $a=0.57$,  $b=0.29$, $b'=1.19$, $b''=1.18$ which gives $E_B=3.4\times 10^{-4}$ at $\Omega/\omega_{10}=0.005$.

Using Eq. (\ref{eq.Eminpryd}) we find the following CNOT truth table error estimates. 
For $^{87}$Rb $124p_{3/2}$ with $\tau=616~\mu\rm s$ we find $\Omega_{\rm opt} = 2\pi\times 16.3~\rm MHz$ ($\Omega_{\rm opt}/\omega_{10}=0.002$) and $E_{\rm cb}=1.3\times 10^{-4}$.
For Cs $112p_{3/2}$ with  $\tau=593~\mu\rm s$ we find $\Omega_{\rm opt} = 2\pi\times 19.5~\rm MHz$ ($\Omega_{\rm opt}/\omega_{10}=0.002$) and $E_{\rm cb}=1.1\times 10^{-4}$.

We see that the truth table error estimates are slightly less than for the lower $n$ situation of Fig. \ref{fig:p_rydlevels}. 
An additional advantage of using higher $n$ states is that the optimal blockade shift is reached with a larger $d\sim 3-4~\mu\rm m$ which may  help to minimize qubit addressing crosstalk in an actual implementation. 
For convenience we have summarized all parameters for the states used in Table \ref{tab.Rydbergsdp}.
\begin{table*}[!t]
\caption{Parameters of the Rydberg states used for QPT simulations. The scaling parameter $b''$ is defined as 
$b''= {\sf B}_{n,n-2}/{\sf B}_{nn}$ for $np_{3/2}, nd_{5/2}$ states and  $b''= {\sf B}_{n,n-1}'/{\sf B}_{nn}$ for $ns_{1/2}$ states.
}
\begin{tabular}{|l|llll|llll|}
\hline
&\multicolumn{4}{|c|}{$^{87}$Rb}&\multicolumn{4}{|c|}{Cs}\tabularnewline
\hline 
Rydberg state &$76p_{3/2}$ &$124p_{3/2}$&$123d_{5/2}$&$82s_{1/2}$&  $70p_{3/2}$& $112p_{3/2}$ &$112d_{5/2}$ &$80s_{1/2}$\tabularnewline
\hline
$\omega_{10}/2\pi$ (GHz)&6.8&6.8&6.8&6.8&9.2&9.2&9.2&9.2\\
$\omega_{n,n-1}/2\pi$ (GHz)&17.0&3.7&3.7&13.7&23.0&5.2&5.1&15.3\\
$\omega_{n,n-2}/2\pi$ (GHz)&&7.5&7.5&&&10.6&10.3&\\
$\omega_{n,n-1}'/2\pi$ (GHz)&&&&-10.4&&&&-8.5\\
$\omega_{n,n-2}'/2\pi$ (GHz)&&&&2.9&&&&6.5\\
$\tau~(\mu\rm s)$&223&616&524&212&211&593&367&191\\
$d~(\mu\rm m)$&1.8&4.5&5.0&2.5&1.4&3.2&3.1&2.2\\
${\sf B}_{nn}/2\pi ~\rm (GHz)$&3.45&2.0&1.9&3.3&4.4&2.6&2.5&3.9\\
$a=\omega_{n,n-1}/\omega_{10}$&2.5&0.55&0.54&2.0&2.5&0.57&0.55&1.7\\
$a'=\omega_{n,n-1}'/\omega_{10}$&&&&-1.5&&&&-0.93\\
$a''=\omega_{n,n-2}'/\omega_{10}$&&&&0.43&&&&0.70\\
$b={\sf B}_{nn}/\omega_{10}$&0.51&0.29&0.27&0.48&0.48&0.29&0.27&0.43\\
$b'={\sf B}_{n,n-1}/{\sf B}_{nn}$&1.1&1.0&0.77&0.95&1.2&1.2&0.98&0.75\\
$b''$&&1.1&0.66&0.08&&1.2&0.95&0.73\\
$b'''={\sf B}_{n,n-2}'/{\sf B}_{nn}$&&&&0.15&&&&0.71\\
\hline
\end{tabular}
\label{tab.Rydbergsdp}
\end{table*}

\subsection{Rydberg $nd_{5/2}$ states}

The  issue of small fine structure splitting of the $p$ states  discussed above, also applies to the $nd_{3/2}, nd_{5/2}$ states.
There are two ways of avoiding this problem. 
As with excitation of the $np$ states  we may assume two-photon excitation from the stretched ground state with $\sigma_+$ light which only couples to $nd_{5/2}, m_j=5/2$ states. Alternatively if the first leg of the excitation is made via the D1 transition ($5s_{1/2}-5p_{1/2}$ in Rb 
or $6s_{1/2}-6p_{1/2}$ in Cs) then only the $nd_{3/2},m_j=3/2$ states can be reached. Since we are setting the separation $d$ to give the desired blockade strength the only difference in the gate error using $nd_{3/2}$ or $nd_{5/2}$ states is due to differences in the lifetime. The lifetimes differ by only a few  percent \cite{Beterov2009} and we will therefore simply consider $nd_{5/2}$ states.  The choice of optimum states and error analysis then follows that in Sec. \ref{sec.Rydbergp}. We have summarized the parameters for the states used in Table \ref{tab.Rydbergsdp}.

\begin{figure}[!t]
\vspace{.3cm}
\begin{centering}
\includegraphics[width=.8\columnwidth]{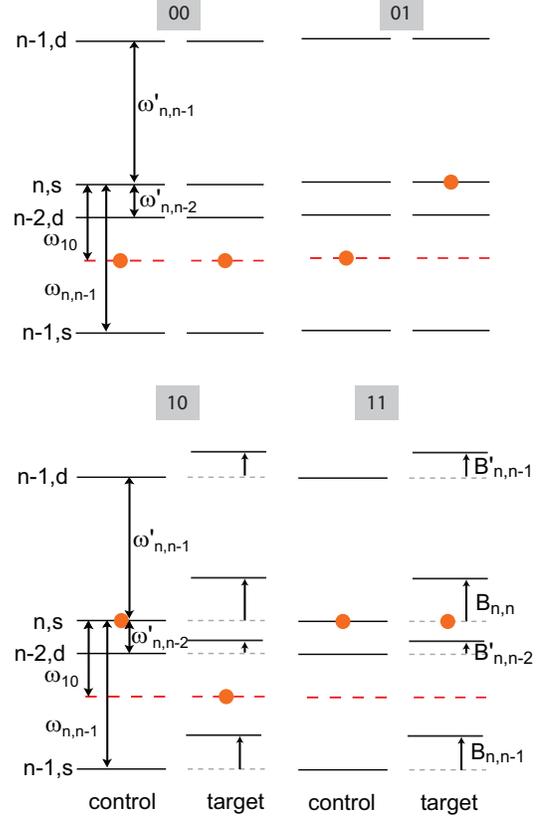}
\par\end{centering}
\caption{(color online) Off-resonant Rydberg excitation for $s$ states with $\omega_{n,n-1}>\omega_{10}$. Primed quantities refer to frequencies or blockade shifts between $s_{1/2}$ and $d_{3/2}$ states. The filled circles indicate the energies where the atoms are excited to with the long-dashed red lines corresponding to excitation of an atom starting in $\ket{0}$.
\label{fig.s_rydlevels} }
\end{figure}

\subsection{Rydberg $ns_{1/2}$ states}

The $ns_{1/2}$ states have no fine-structure but there is an additional complication since it is not possible to use angular momentum selection rules to couple to $ns_{1/2}$ states, but not $nd_{3/2}$ or $nd_{5/2}$ states. We must therefore consider off-resonant coupling to additional Rydberg levels. The situation for $\omega_{n,n-1}>\omega_{10}$ is shown in Fig. \ref{fig.s_rydlevels}. The leading contributions to blockade errors for the computational basis states are
\begin{widetext}
\begin{subequations}
\begin{eqnarray}
E_{00} &=& \frac{3}{2}\left[\frac{\Omega^2}{\omega_{10}^2}+\frac{\Omega'^2}{(\omega_{10}-\omega_{n,n-2}')^2}+ \frac{\Omega^2}{(\omega_{n,n-1}-\omega_{10})^2}\right] \\
E_{01} &=&  \frac{2}{3}E_{00}+ \frac{\Omega'^2}{2\omega_{n,n-2}'^2}\\
E_{10} &=& \frac{\Omega'^2}{\omega_{n,n-2}'^2}+\frac{\Omega^2}{2(\omega_{n,n-1}-{\sf B}_{n,n-1}-\omega_{10})^2} + \frac{\Omega'^2}{2(\omega_{10}-\omega_{n,n-2}'+{\sf B}_{n,n-2}')^2}\\
E_{11} &=& \frac{\Omega'^2}{\omega_{n,n-2}'^2}+\frac{\Omega^2}{2{\sf B}_{n,n}^2} + \frac{\Omega'^2}{2(\omega_{n,n-2}'-{\sf B}_{n,n-2}')^2}.
\end{eqnarray}
\label{eq.rydserror1}
\end{subequations}
\end{widetext}
Here $\omega', {\sf B}'$ refer to frequencies and couplings between $s_{1/2}$ and $d_{3/2}$ states and $\Omega'$ is the Rabi frequency for $nd_{3/2}$ excitation via the D1 transition from the ground state so that we need only consider $nd_{3/2}$ states. 
We introduce dimensionless parameters $a=\omega_{n,n-1}/\omega_{10}$, $a'=\omega_{n,n-1}'/\omega_{10}$, $a''=\omega_{n,n-2}'/\omega_{10}$,
$b={\sf B}_{n,n}/\omega_{10}$, $b'={\sf B}_{n,n-1}/{\sf B}_{n,n}$, $b''={\sf B}_{n,n-1}'/{\sf B}_{n,n}$, $b'''={\sf B}_{n,n-2}'/{\sf B}_{n,n},$ and $c'=\Omega'/\Omega.$ Using $\sigma_+, \sigma_-$ polarized excitation light for the ground-D1 and D1-Rydberg transitions $c'= 1.31$.
Since $|\omega_{n,n-1}'|\gg |\omega_{n,n-2}'|$ we have neglected terms due to  coupling to $n-1,d_{3/2}$ in (\ref{eq.rydserror1}). Additional checks with this level included give not more than 5\% increase in the error averaged over the computational states. Since the computational cost of adding an additional Rydberg level in the master equation simulations is large we have neglected this small correction and performed master equation  simulations with the three Rydberg states, $|r\rangle=|n,s_{1/2}\rangle, |r_1\rangle=|n-1,s_{1/2}\rangle, |r_2\rangle=|n-2,d_{3/2}\rangle$.  

Numerical checks using Eqs. (\ref{eq.rydserror1},\ref{eq.Eminpryd})  were used to select the states listed in Table \ref{tab.Rydbergsdp}.

\end{document}